# Autoencoding sensory substitution

**Viktor Tóth**

**School of Science**

Thesis submitted for examination for the degree of Master of Science in Technology.

Espoo 28.02.2019

**Thesis supervisor:**

Prof. Lauri Parkkonen

**Aalto University**
School of Science

Author: Viktor Tóth

Title: Autoencoding sensory substitution

Date: 28.02.2019      Language: English      Number of pages: 8+96

Department of Neuroscience and Biomedical Engineering

Professorship: Biomedical Engineering

Supervisor and advisor: Prof. Lauri Parkkonen

Tens of millions of people live blind, and their number is ever increasing. Visual-to-auditory sensory substitution (SS) encompasses a family of cheap, generic solutions to assist the visually impaired by conveying visual information through sound. The required SS training is lengthy: months of effort is necessary to reach a practical level of adaptation. There are two reasons for the tedious training process: the elongated substituting audio signal, and the disregard for the compressive characteristics of the human hearing system.

To overcome these obstacles, we developed a novel class of SS methods, by training deep recurrent autoencoders for image-to-sound conversion. We successfully trained deep learning models on different datasets to execute visual-to-auditory stimulus conversion. By constraining the visual space, we demonstrated the viability of shortened substituting audio signals, while proposing mechanisms, such as the integration of computational hearing models, to optimally convey visual features in the substituting stimulus as perceptually discernible auditory components. We tested our approach in two separate cases. In the first experiment, the author went blindfolded for 5 days, while performing SS training on hand posture discrimination. The second experiment assessed the accuracy of reaching movements towards objects on a table. In both test cases, above-chance-level accuracy was attained after a few hours of training.

Our novel SS architecture broadens the horizon of rehabilitation methods engineered for the visually impaired. Further improvements on the proposed model shall yield hastened rehabilitation of the blind and a wider adaptation of SS devices as a consequence.

# Acknowledgments

I would like to express my very great appreciation to my friends in Otaniemi for their companionship in the times of darkness and beyond. I would further like to offer my special thanks to Viljar Rúnarsson, who provided unconditional assistance, and above all, tolerated my blindfolded alter ego.

Ajka, 23.11.2018

Viktor Tóth

# Contents

# Abbreviations

| | |
|---|---|
| AC | auditory cortex |
| AM | amplitude modulation |
| AN | auditory nerve |
| AEV2A | Autoencoded Vision to Audition |
| BM | basilar membrane |
| CARFAC | cascade of asymmetric resonators with fast-acting compression |
| CB | congenitally blind |
| CNN | convolutional neural network |
| CF | center frequency |
| CORF | combination of receptive fields |
| DNN | deep neural network |
| DRAW | deep recurrent attentive writer |
| EB | early blind |
| FIR | finite impulse response |
| FM | frequency modulation |
| GAN | generative adverserial networks |
| HRTF | head-related transfer function |
| IC | inferior colliculus |
| IIR | infinite impulse response |
| ILD | interaural level difference |
| ITD | interaural time difference |
| jnd | just-noticeable difference |
| LB | late blind |
| LGN | lateral geniculate nucleus |
| LOC | lateral occipital cortex |
| LOtv | lateral-occipital tactile–visual |
| LSO | lateral superior olive |
| LSTM | long short-term memory |
| ME | mixture of experts |
| MFCCs | mel-frequency cepstral coefficients |
| MSO | medial superior olive |
| SM | spatial modulation |
| SPL | sound pressure level |
| SS | sensory substitution |
| TV | The vOICe |
| V2A | visual-to-auditory, vision-to-audition |
| VAE | variational autoencoder |

# 1 Introduction

> I had lost my sight, but I got something back in return. My remaining four senses functioned with superhuman sharpness. But most amazing of all, my sense of sound gave off a kind of radar sense. (Daredevil, 2003)

36 million people live blind, out of which 1.4 million are children [1]. Due to population growth and aging, these numbers are ever increasing, potentially reaching 115 million by 2050 [2]. While prevention and cure exist for most visually impaired, such as replacement of lens for patients with cataracts, availability of such treatments in underdeveloped countries is scarce. Furthermore, certain blinding diseases, like advanced glaucoma, still lack remedy, while recent invasive advances in treatment of retinal diseases are complicated, expensive and provide limited restoration of vision [3].

Sensory substitution (SS) devices are designed to supplement sensory information of the impaired modality through another, active sense in a non-invasive manner. In case of visual-to-auditory (V2A) SS devices, the visual information is translated to audio, yielding sight via hearing. In practice, the blind may wear a camera on the head, the images taken are translated one-by-one into soundscapes and played in earphones. V2A SS devices have been used extensively in research about cross-modal plasticity, metamodality [4] and synaesthesia [5] research, but the devices have seen only incremental development since 1992. These improvements addressed the encoding of color into audio [6], the ability to zoom in on the image translated [7], or the application of interaural disparities [8]; however, none has challenged the V2A encoding logic of The vOICe [9], and achieved substantially better results in vision restoration.

Adaptation to SS devices is inherently a compromise where the substituting (hearing) modality is partially traded for the substituted (vision). The required training is lengthy, the effort necessary to reach a practical level in the employment of these devices is measured in months. Reaching an ideal synaesthetic experience, where a sound is automatically perceived as a visual stimulus, may take 5 years for the late-blind to achieve [10].

One of the the major technical reasons behind the required lengthy training is the slow V2A conversion logic used so far. V2A conversion algorithms take an image as input and produce a soundscape as output. There is a trade-off between the length of the soundscape and the loss of information the sound is to convey [11]: the longer the soundscape is, the more detailed is the visual information one can extract. On the other hand, increased length means increased delay between reality and the perception of the soundscape, which interferes with multisensory integration that occurs on a relatively short timescale [12]. Multisensory integration and cross-modal learning are essential to perceive the soundscapes in an intuitive manner [13]. For instance, the blind best trains on V2A SS devices by integrating tactile information and the presented soundscapes. Ultimately, we want the shortest possible sound representation of every conceivable image without the loss of information before being projected on the cortex via the auditory pathway. The methods applied so far reach an upper limit in this respect [11], which calls for a new perspective.

Previous V2A SS device designs have taken advantage of behavioral studies of multisensory integration, but barely incorporated the study of acoustics and human hearing models [14], brain imaging studies of sensory coding and, research in cross-modal plasticity and metamodality [4]. Acoustics and the study of sensory coding provide computational models and neural imaging results, respectively, that define implicitly [15–19] and explicitly [20–26] the subspace of soundscapes that the human auditory system can code for and would not compress away. Utilizing such subspaces of soundscapes to represent visual information in a V2A encoding scheme, we ensure that the auditory system can differentiate between the sounds corresponding to different visual information.

The study of cross-modal plasticity and metamodality outlines the framework of how the received auditory information is transferred to and processed in other sensory areas. By understanding more about the cross-modal connections and the receiving modality, we can specifically tailor the soundscapes so they are easily mapped onto the visual cortex, reducing the amount of necessary cross-modal plasticity [27].

The "Holy Grail" [28] of V2A SS is the synaesthetic experience: visual perception consistently and involuntarily conjured solely by sound. Synaesthesia of this kind may be achieved by unmasking the previously mentioned cross-modal connections and by further building neural links between the modalities [29]. Synaesthesia can be induced by the use of psychedelic drugs [30], including psilocybin, which has been shown to increase the excitability of the neural networks bridging between sensory regions [31].

The V2A conversion logic in previous SS devices is explicitly defined, i.e. every visual feature, or every pixel has a well-specified linear contribution to the constructed audio signal, dependent only on the pixel's position and shading or color content. On one hand, such a conversion logic can be clearly explained in plain words, which facilitates SS learning [32,33]. On the other hand, it fails to recognize the distribution of plausible (natural) images and treats the encoding of all possible pixel constellations uniformly. It reserves acoustic codes for images that are unlikely to occur, which is part of the reason such V2A conversion methods hit the previously mentioned lower limit in soundscape length. Deep neural networks applied in computer vision admit to the prior distribution of images, similarly to our visual cortex, resulting in comparable effectiveness in encoding likely visual representations in orders of magnitude lower dimensional space than the original picture [34].

The major contribution of this study is a novel, implicit V2A conversion method molded into a deep neural network that fuses the compression mechanisms of human hearing, the neural coding of the auditory and visual cortices, and the characteristics of cross-modal learning and multisensory integration. The deep learning model is realized as a variational recurrent autoencoder [35], including a sound stream synthesizer and a simplified computational model of human hearing, virtually drawing the input image on the primary visual cortex in an iterative manner.

The efficacy of this model is demonstrated in two experiments. In the first experiment, the author went blindfolded for 5 days while performing SS training on hand posture discrimination. The second inquiry assessed the accuracy of reaching movements given objects on a table. We successfully trained the deep learning models on different datasets to execute V2A stimulus conversion. By constraining the visual space, we demonstrated the viability of shortened substituting audio signals, while proposing mechanisms, like the integration of computational hearing models, to preserve the substituted visual features in the substituting stimulus as perceptually discernible auditory components. In both case studies, above chance level accuracy was attained after a few hours of training.

# 2 Background

## 2.1 Sensory coding

Understanding the coding mechanism of both the substituting (auditory) and substituted (visual) sensory areas is essential in the design of SS devices. First, we need to establish the set of soundscapes that humans can distinguish. Then we examine the ways in which such sound stimuli are encoded in the auditory system and cross-modally in the visual cortex.

### 2.1.1 Substituting modality

The temporal and spectral accuracy of stimulus coding in the auditory pathway should define the SS stimulus presented to it, due to the compression and the nonlinearities the signal suffers along the pathway. When two tones of similar frequencies are played simultaneously, they asymmetrically mask each other in the cochlea. The higher frequency tone tend to be masked by the lower one. Depending on the sound intensity, the higher frequency tone can completely disappear perceptually [36]. If we fail to consider this compression mechanism, and encode visual cues in simultaneously presented sounds of similar pitch, the signal deteriorates before even reaching the cortex. The inner hair cells translate the continuous signal of cochlear vibration into discrete action potentials. This quantization further reduces information, discounting loudness levels between the quantiles. Moreover, if visual details are rendered in loudness levels, we need to regard for the smallest increment of perceivable amplitude, the just-noticeable difference (jnd).

### 2.1.2 Cross-modality

The cerebral cortex is a surface of 2200–2400 cm$^2$, with a thickness of 2.5–3.0 mm [37]; it can be considered as a two-dimensional sheet of 20 billion neurons, organized in cortical columns having lateral, feed-forward, and feed-back connections [38]. Once the visual cues are projected on the sheet via the auditory pathway in a V2A SS scenario, cortico–cortical connections are available to be established under cross-modal plasticity [39]. In part, this study builds upon the following assumption: if the nature of the sound-encoded visual information is similar to the archetypal visual features coded for in the occipital cortex, the cross-modal plasticity may be sped up. The similarity in the nature of visual information here refers to the amount of coding transformation necessary to map the auditory coding of the SS stimulus to the corresponding occipital stimulation reflecting the encoded visual cues as it would in a sighted person's cortex. The less transformation is necessitated between the auditory and visual coding schemes, the less neural plasticity is required to consolidate the cross-modal connections. Provided that this assumption is correct, and that the striate cortex largely contributes to the V2A SS stimulus processing [40–42], understanding of the substituted sensory coding becomes critical.

In summary, if we aim to better SS devices, we need to formulate the sensory coding of both substituting and substituted modalities, focusing on the limitations of the former and the internal neural encoding of the latter. This is especially true for V2A SS, where the auditory pathway acts as a bottleneck of visual information transfer, due to encompassing only around 30,000 fibers, while the optic nerve consists over 1 million fibers [43].

### 2.1.3 Substituted modality

The Holy Grail of SS is likely to emerge from the excitation of the occipital cortex via cortico–cortical connections rooted at the primary auditory cortex [10, 39, 44]. In this study, we only examine early sensory processing of the modalities. More specifically, the auditory pathway is reviewed until and including the primary auditory cortex, while only the primary visual cortex is studied within the visual pathway. These sensory areas bring the most relevance in V2A SS for the following reasons: 1) cross-modal plasticity appears to be maximal at the most modality-specific brain regions, like the early visual cortex, in absence of input, they become available for auditory processing [45, 46], 2) ipsilateral, direct and indirect A1 to V1 cross-modal plasticity is the most apparent [39, 47], rendering thalamo–cortical SS connections doubtful, 3) the optic chiasm and the optic radiation of the blind suffers from atrophy due to the lack of visual input [48], which renders pre-V1 areas unavailable for SS processing. Hence, in the following we solely consider the encoding of the early auditory pathway and the primary visual cortex. In the case of SS device design, this choice has practical reasons as well: the coding strategy of higher sensory areas are less documented than early regions. Thus, tailoring the SS stimuli to complex cells instead of simple cells of V1 would cause substantial difficulty.

### 2.1.4 Information theory in neural coding

There are two types of neural coding strategies: rate coding and temporal coding [49]. Neurons encoding information in the number of spikes fired under a time window said to perform rate coding. On the other hand, neurons execute temporal coding, when they encode information mainly in the timing of emitted spikes. We see examples of both strategies and their mixture throughout the cortex. For instance, V1 cells fire more spikes in bursts when the orientation they code for fits the bar angle in their receptive field. Temporal coding is frequent in the auditory pathway; in the auditory nerve, neurons phase lock to the auditory stimulus, firing action potentials in a rhythm, when the amplitude of the stimulus is at its positive peak [37].

In order to statistically model rate codes, we first assume that the spikes are generated by a random Poisson process, which only requires the expected rate of fire as a parameter. The noisier the neuron, the higher this lambda parameter is. After a set of stimuli is given as input to the process, we can apply ANOVA to find relevant variables the neuron seems to code for [49]. When a variable, e.g. loudness, is tracked down we assess the rate of spikes as the function of sound level. By plotting this rate along the sound level of given stimulus, we arrive at a tuning curve of the neuron. In essence, the tuning curve shows the distribution of firing rate marginalized for a certain variable.

When statistically modeling temporal codes, we tend to compute the mutual information of ever shrinking windows of neural activity, given the stimulus, until no more significant information is gained. Once the proper window size is attained, the turning curve is computed akin to rate coding.

Information theory provides us simple, straightforward methods to quantify the ability of neural assembles to code for a stimulus [50]. The amount of information, or the amount of uncertainty, can be defined for both the stimulus and the response: $H[\Theta] = -\sum_{\theta} p(\theta) \log_2 p(\theta)$. Similarly, the amount of information is derived considering both the stimulus and the response: $H(s, r) = -\sum_{s,r} p(s, r) \log_2 p(s, r)$. Mutual information is then specified as: $I(s, r) = H(s) + H(r) - H(s, r)$. If the neural assembly in question codes for the stimulus, then $I(s, r) > 0$, because the uncertainty is decreased once the stimulus and response are combined. Small enough $I(s, r) - H(s)$ implies that the stimulus is reliably encoded in the neural population.

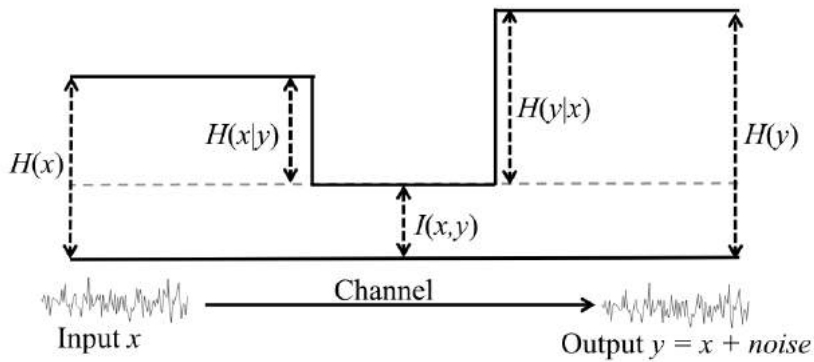

Figure 1: Relations between quantities of information theory. Reprinted from [51].

Theoretically, if we were able to derive the mutual information between any arbitrary sound and the auditory cortex (AC) response, we could select those soundscapes that are properly neurally encoded. As the current brain imaging techniques and computational neural models are not yet developed enough to achieve this, the function of mutual information by stimulus has to be approximated. In this paper, approximations are extracted from neuroimaging, psychoacoustics and psychophysical research.

In conclusion, information theory provides a theoretical framework, in which one could map out the subspace of stimulus that is reliably encoded in the cortex. Such stimuli then may be used in a SS setting as an input to the substituting modality. In case of V2A SS, a subset of soundscapes can be found that are accurately coded in AC. Then, a one-to-one V2A function is specified that takes an image as input and returns one of previously defined soundscapes. Hence, we can be ensured that the visual information is properly encoded in AC.

## 2.2 Visual coding

### 2.2.1 Functional organization

In the visual cortex of the sighted, features of motion, color, and surface texture are encoded [52]. Input coming from the lateral geniculate nucleus (LGN) is processed in a parallel fashion [53, 54]. The surface of the primary visual cortex (V1) is laid out in a retinotopic manner: the neurons tuned to the fixation point are positioned at the posterior-ventral end of the cortex, while the cells responding to peripheral visual information are arranged more anterior. A single neuron in V1 has a compact receptive field. Neurons are grouped into cortical columns [55], some of which are ocular dominant, preferring visual input from one of the eyes only [56]. Ocular dominant columns include orientation columns, which are tuned to contrast edges, or bars of a given orientation. These columns of orientation specific neurons are repeated in V1 $\pi$ periodically, with patches of cells reacting to similar angles being adjacent to each other. Orientation columns are intertwined with local intracortical connections and long-range lateral connections to cells of similar orientation tuning [57].

From V1, two main streams arise: the ventral and the dorsal stream. The ventral stream, or the "what" pathway, processes object vision [58]. The dorsal stream, or the "where pathway", is mainly responsible for coding the spatial location of objects [59].

Retinal ganglion cells and similarly cells in LGN are excited by circular center-surround receptive fields. We differentiate between two types of ganglion cells: on–off and off–on cells. On–off cells are tuned to relatively high luminance in the center, surrounded by relatively low luminance, while off-on cells code for the opposite. V1 simple cells process this incoming information by stitching together the adjacent center–surround receptive fields, responding to elongated regions of alternating high and low luminance, that is, a contrast bar or edge [60]. Orientation of the edge is rate coded, while the amount of contrast is related to the latency to the response onset [49, 61].

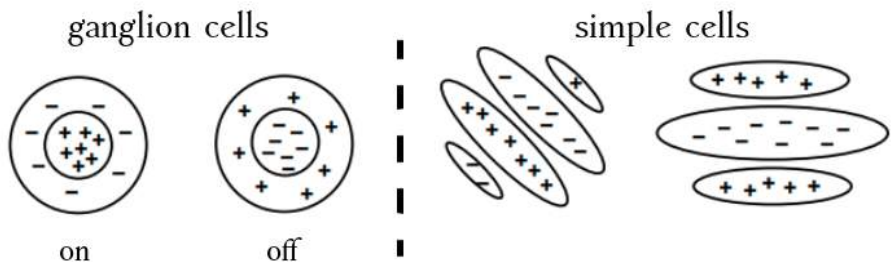

Figure 2: Receptive fields of ganglion and simple cells. Adapted from [62].

Comparison of the visual cortex of sighted and blind individuals poses difficulty, due to the recent lack, or complete absence of visual experience in the case of late and congenitally blind, respectively. Cross-modal plasticity is then driven by this lack of input to establish functional connections with other sensory brain regions, which in turn, influences the structure of the occipital cortex [39].

In the work of Striem-Amit and colleagues [63], resting-state functional connectivity of the visual cortex of blind and sighted is compared to assess the extent of structural difference. More specifically, the large-scale segregation of visual functional networks is examined on the basis of eccentricity, laterality and elevation. In that study, the eccentricity, laterality and elevation maps of blind and sighted were grouped one-by-one by applying the k-means clustering algorithm. When clustered, the same brain regions were grouped together according to the different maps, regardless of whether the map was extracted from a sighted or a blind subject. This result indicates that the large-scale structure of the retinotopic network in the visual cortex is not dependent on visual experience. Striem-Amit and others further show that the eccentricity bias inherent in V1 connections to the fusiform face area (FFA) and the parahippocampal place area (PPA) is equivalent to the bias present in the functional connections of the sighted; central V1 having stronger connections to FFA, while peripheral V1 neuron activations coinciding more with PPA activity [63].

The development of the visual cortex seems to be influenced by experience dependent visual radiation, activity-independent molecular factors and activity-dependent effects as well as spontaneous retinal waves [64]. The findings of Striem-Amit and colleagues [63] do not imply that the more micro-scale functional organization of V1 is identical regardless of visual experience. Hence, orientation selective cells might not exist in the blinds' V1.

Beyond the primary visual cortex, the macro-scale organization of the visual pathway develops into the ventral and dorsal streams, even in the complete absence of visual experience [65].

In the design of V2A SS devices, we need to translate visual information to sound. Technically, the visual features may be the pixels of an image or more abstract properties, such as edges. Edges of an image requires substantially less coding space than the shade or color information. This is appealing for a branch of SS where the substituting modality harbors orders of magnitude less bandwidth than the substituted.

### 2.2.2 Computational models

Edge detection is the computational image processing equivalent of the bar orientation coding performed in the primary visual cortex. To extract edges from images, either biologically inspired models or simpler convolutional functions are employed; we begin to describe the latter.

Several convolutional methods exist, essentially varying in their filter shape. The Sobel operator [66] applies two filters with a relatively strong contrast at the center: Gx and Gy. Gx responds to vertical while Gy to horizontal edges when convolved. The strength of an edge regardless of its orientation is given by $\sqrt{G_x^2 + G_y^2}$; the orientation is computed as $\arctan(G_y/G_x)$.

The Canny edge detector takes the intensity gradient as an input, e.g. the output of the Sobel operator, sharpens the already detected edges by employing non-maximum suppression, and further selecting the dominant and adjacent-to-dominant edges by applying hysteresis thresholding [67].

The Gabor filter, named after the exceptional Hungarian physicist Dénes Gábor, is a Gaussian kernel function modulated by a sinusoidal plane wave [68, 69]. When applied to edge detection, a filterbank is constructed consisting of Gabor filters with different orientation and scale. Mehrotra and colleagues [69] demonstrated that receptive fields in the cat's visual cortex resemble Gabor filter shapes.

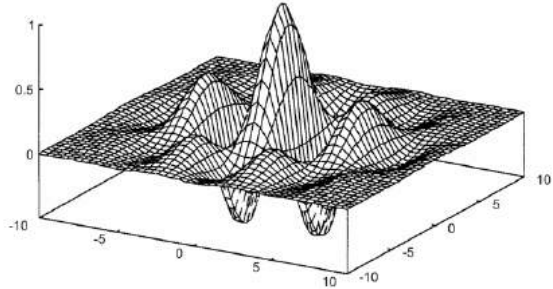

Figure 3: Two-dimensional Gabor filter.

Biologically-inspired edge detection models incorporate the early hierarchy of the visual system, including retinal ganglion cells, LGN and V1 simple cells. Azzopardi and others [70] argues that the Gabor filter and other convolutional approaches ignore the functionality of LGN neurons and fail to emulate simple cell properties, such as cross-orientation suppression, response saturation and contrast-invariant orientation tuning. Cross-orientation suppression stands for the lateral inhibition applied to neurons in the same receptive field, but with different orientation; such inhibition aids to diminish weaker edges adjacent to a strong one. Response saturation is embodied in the sigmoid shaped response function of V1 cells as a function of contrast intensity. Azzopardi and colleagues [70] developed an edge detection method that integrates the previously mentioned properties, called the Combination of Receptive Fields (CORF) model. CORF defines LGN receptive fields as a pair of overlayed Gaussian patches, one smaller, one wider with opposite signs to mirror the on-off and off-on functionality.

LGN cells are bundled into sub-units, representing the response of a dendrite branch that simple cells integrate. Sub-units receive input from LGN neurons aligned in two line segments, on-off cells in one and off-on cells in a parallel nearby segment. The contrast edges these sub-units react to are then combined in a weighted sum by simple cells. Similar to Canny, CORF further performs non-maxima suppression followed by hysteresis thresholding on the V1 cell responses.

Azzopardi and colleagues [71] improved the CORF model, introducing a push-pull mechanism causing the inhibition of cells in the same receptive field with opposite contrast. Compared to the previous version, push-pull CORF is biologically more viable, and demonstrates more reliable edge detection of noisy images.

In conclusion, the push-pull CORF approach yields the highest signal-to-noise ratio in extracting the relevant edges. However, the current CORF implementation hardly applies parallel computation [72], scaling badly as the input image resolution rises. Parallel implementations of image convolution algorithms exist and runs Gabor and Sobel filters an order of magnitude faster. Among convolutional edge detectors,

Gabor filters emulate simple cell response patterns the most accurately.

## 2.3 Auditory coding

### 2.3.1 Functional organization

Sound is ultimately defined by pressure changes, or vibrations in a given medium. The auditory system has evolved to detect subtle pressure pulses in the environment and also to integrate such pulses into the identification of loudness, pitch, spatial position and the modulation of these variables over time. Anything that we perceive can be boiled down to 50,000 pressure values a second [50], which bandwidth is substantially compressed by the auditory system, before the sound radiates into our consciousness.

The pressure wave enters the auditory periphery through the external auditory canal and causes the vibration of the eardrum, then the ossicles, which, in turn, induces wave propagation in the basilar membrane (BM). In the cochlea, the BM reacts to the sound in a tonotopical fashion. The sites along the BM respond to distinct distributions of frequencies: basal sites respond to higher while apical sites to lower. The generated traveling wave on the BM then depolarizes hair cells. By releasing neurotransmitters, the inner hair cells induce action potentials in the auditory nerve (AN). Sequence of action potentials of the AN transfers auditory information to the cortex [73]. Compared to the 1 million fibers of the optic nerve, the AN only contains around 30,000 [43]. This shows the bandwidth gap between the two modalities and indicates the difficulty of V2A SS design. As far as the information of pressure amplitudes is propagated to the auditory periphery, the following rule applies: high-intensity sound drives larger amplitude vibrations on the BM, which causes higher depolarization of the inner hair cells, which responds by releasing more neurotransmitters that induces higher firing rates in more AN axons [50]. Outer hair cell responses are shaped by the sound wave and serve as a feedback mechanism to amplify certain frequencies.

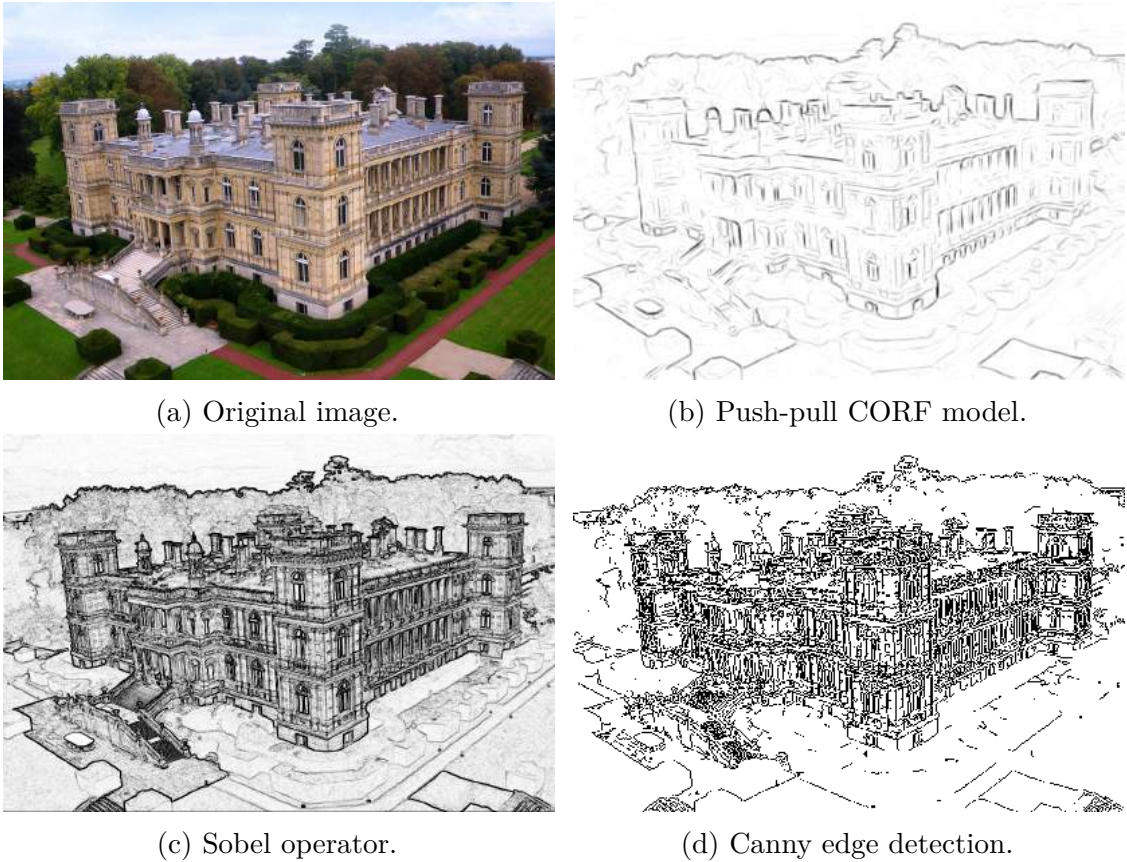

(a) Original image.

(b) Push-pull CORF model.

(c) Sobel operator.

(d) Canny edge detection.

Figure 4: The output of three edge detection algorithms and the input image. Sobel and Canny representations are more detailed, but they contain an abundance of unnecessary information.

The cochlear nuclei constitutes as the first neural processing stage. Here, the auditory system divides into the dorsal, what, and ventral, where, pathways, akin to the visual system [74]. The ventral pathway performs binaural fusion in the superior olivary complex. Within the superior olivary complex, the medial superior olive (MSO) computes the azimuth of the sound by measuring interaural time differences (ITD) contralaterally. Ipsilateral delay lines and varying contralateral axon lengths cause the binaural signals to reach the MSO at different times, where the bilateral sound information is essentially cross-correlated [14]. The lateral superior olive (LSO) detects interaural level differences (ILD), and computes the azimuth position of the sound source at higher frequencies [21].

Neural activity from dorsal and ventral pathways meet at the inferior colliculus (IC). IC is responsible for detecting amplitude and frequency modulation, and it also integrates the results from LSO and MSO representing spatial position, at this stage, independent from the frequency of the sound. IC projects its output to the medial geniculate nucleus (MGN), which functions as a relay between IC and AC. While still maintaining a tonotopy map already present in AN, AC further processes harmonic pitch, indicative of its long-range lateral connections [75]. Moreover, AC codes for auditory space, frequency and amplitude modulation [20], and auditory entities [22, 23].

### 2.3.2 Loudness

In general, we can postulate that high sound intensity drives larger amplitude vibrations of the BM, which causes higher depolarization of the inner hair cells, which responds by releasing more neurotransmitters that induces higher firing rates in more AN axons [50]. The inner hair cells, receiving the tonotopic input from the BM, function as half-wave rectifiers and only propagate the positive half of the signal [14].

Neurons in the AN encode sound intensity by the combination of rate and phase coding. The firing rate, in general, increases by the increase of sound pressure level (SPL), measured in decibels. Between 30 and 50 dB, the spiking rate in the AN correlates well with SPL. The threshold of hearing is at 0 dB, while the pain threshold is around 130 dB, though sound stimuli above 70 dB are considered annoying. High spontaneous rate fibers, about 75% of all rate fibers, encode weak signals up to 50 dB SPL, and 25% of medium and low spontaneous fibers respond to more intense soundwaves [50]. At higher sound levels, the fibers begin to spread the excitation to fibers that supposed to have a CF even an octave away. In other words, the tuning curve of AN fibers widens as the signal level rises [73]. At high SPLs, when place coding of pitch is disturbed, phase locking comes to the rescue: nerve fibers can skip soundwave cycles to convey relatively lower sound intensity in the high SPL regime.

There is a logarithmic relationship between sound intensity and the perceived loudness. However, logarithmic functions harbor singularity at 0, which would translate to an unlimited capacity to perceive arbitrary small sound level ratios [14]. Stevens solved this theoretical problem by substituting the logarithm with a power function [76]: $f(x) = x^\alpha$, where $\alpha = 0.3$ is set to accommodate the human auditory system [77]. The same exponent is applied in the study of Młynarski and others [78] for instance, after the Hilbert envelope of the signal is taken to emulate cochlear amplitude compression. In general, we can differentiate between 240 or so intensity levels, corresponding to 0.5 dB steps between them [50]. However, the actual mapping of intensity to loudness in the human auditory system is more complicated than this power relationship. For instance, perception of loudness depends on the frequency of the sound [14]. The famous Fletcher–Munson curves define the equal-loudness contours across frequencies [79].

As the curves show in Figure 5, sound levels between 600 and 2000 Hz are perceived about the same, but around 3 and 4 kHz, hearing becomes more sensitive. The auditory system is quite negligent to intensity differences below 300 Hz. The ISO 226:2003 standard [80] describes the latest version of the Fletcher-Munson curves.

By understanding the extent to which sound levels are encoded in the human auditory system, we can scale the loudness of V2A SS auditory stimuli, so that the information stored in the level is not lost before reaching the cortex. Furthermore, we can design the SS stimuli to encode more information in the loudness of sounds with frequencies, to which the auditory system is more sensitive. Finally, we need to make sure that our SS model does not rely on the negative half of the audio signal much, as it gets cut off once it is translated to neural spikes.

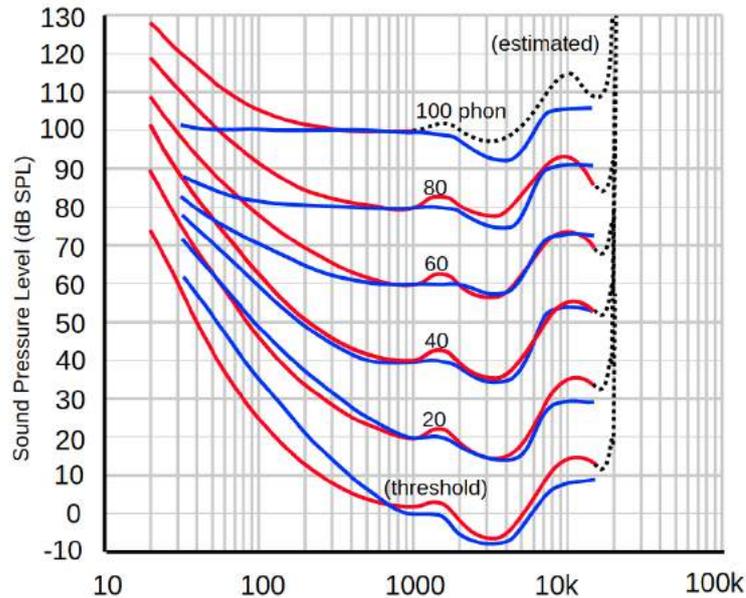

Figure 5: Fletcher-Munson equal-loudness contours shown in blue, the latest ISO 226:2003 revision in read. The horizontal axis represents sound frequency in Hz.

### 2.3.3 Tonotopy

In this study, we refer to tone as an umbrella term for pitch and timbre. Pitch is a psychoacoustical, subjective attribute of sound; it represents the perceived frequency. Historically, pitch was considered the main frequency component of a sound, derived by the means of Fourier transformation. However, we can construct audio samples, which are perceived at a certain pitch, even though the Fourier decomposition does not contain the corresponding frequency: e.g. iterated ripple noise [81]. Thus, pitch is not only a frequency, but also a time-domain property, describing the rate of repetition, or the inverse of the temporal delay between prominent spikes [82]. The misunderstanding of equalizing pitch and frequency stems from the fact that they represent the same property in pure sinusoids, which were extensively, if not exclusively, applied in auditory research in the past [14]. When mixing filtered noises and non-harmonically related sinusoids for instance, pitch cannot be defined by the power spectrum alone; understanding of cochlear mechanics and auditory neural coding principles is a necessity.

Timbre is the difference between two distinct sounds of the same pitch and loudness. It is approximately described by the spectrum and envelope of the sound, which in turn are influenced by the amplitudes and phases of harmonics found in the tone. Harmonics are based on small integer ratios of frequency; if the fundamental frequency is said to be $f_0$, then the harmonics are $2f_0, 3f_0, 4f_0$ ..., all being periodic in the fundamental frequency. Encoding timbre is essential, for instance, to differentiate between environmental sounds (lack of harmonics) and speech (rich in harmonic structure); humans are capable of resolving 5 to 8 harmonics of a complex tone [75]. A fundamental frequency combined with its harmonics is observed as a single sound, while non-harmonic pure tones are detected separately.

Coding for tone starts in the cochlea, where different sites of BM respond primarily to different frequencies of sound. Each site is generally modeled as a bandpass filter with a center frequency (CF) that it responds to the most. A cascaded filterbank of such bandpass filters sufficiently approximates the underlying mechanics [14]. Tonotopy is decomposed in the cochlea through these filters. The decomposed signal is further propagated in AN, where different frequencies are encoded in different neurons, each having a CF. This tonotopic map is conserved until AC [37, 65].

Auditory neurons tend to phase lock to the stimulus, firing in bursts at a certain phase of the signal. Cells spike at positive phases, as the auditory neurons only code for the positive half of the auditory stimulus. Volley theory [The Perception of Low Tones and the Resonance] further states that by reacting to every n'th phase, groups of auditory neurons can code for frequencies as high as 5 kHz, while pure phase coding could only account for 500 Hz and below.

A small portion of neurons (less than 20%), beginning at the cochlear nucleus, responds to harmonics. Some neural cells in A1 even react to pairs of tones presented with a delay of 120 ms, and frequency separation of about 1 octave [75].

Although the primary AC still has a tonotopic organization [23], it predominantly encodes more abstract auditory entities than pitch or even spectro-temporal patterns [22]. A1 neurons barely respond to sustained single-frequency pure tones or noise stimulus; they mostly react to stimulus onset, while later neural activity is less regular [50, 83]. Such a coding strategy is reflective of the composition of natural signals, like vocalizations, where pitch and timbre transitions are fast. It was also demonstrated that activations in the primary AC correlate more strongly to ethologically relevant abstract categories of sound [22].

The tonotopic map follows a logarithmic distribution of CFs: $f_C = A(10^{\alpha x} - k)$, where $f_C$ is in kHz and $x$ is defined as the distance from the apex of the BM, from 0 to 1. The constant $A$ determines the range of CFs, $k$ varies across species, but resides dominantly between 0.8 and 1, while $\alpha$ is consistently 2.1 among mammals [77, 84]. The logarithmic distribution of frequency-encoding filters determine human's ability to discriminate between pitches, allowing them to perceive frequency ratios rather than frequency differences. So, while we can discern pure sinusoids of 420 and 421 Hz, we can hardly tell the difference between 8000 and 8050 Hz. Sun and others [83] applied linear discriminant analysis, among other methods, to infer the frequency of the sound stimulus from single-neuron recordings of marmoset monkeys. They suggested that 10–20 ms of A1 recordings post stimulus onset should suffice to classify sound frequency, which indicates an upper bound on the amount of necessary stimulus interval, while the Nyquist-Shannon sampling theorem provides a lower bound.

Beside the logarithmic frequency compression, the cochlea harbors some obscene nonlinearities. First, the interaction of two or more waveforms propagating along the BM causes intermodulation distortion. The frequency of the emerging distortion products [85] take the form $f_2 - f_1$, $(n+1)f_1 - nf_2$, $(n+1)f_2 - nf_1$, where $n = 1, 2, 3, ...$ [73]. Second, suppression, or masking, occurs when two or more tones of close-by frequencies are processed by the cochlea simultaneously. The magnitude of suppression is the highest when the probe tone is close to the CF of the suppressor tone [77]. We also see suppression thresholds decrease at lower frequencies and increase with higher suppressor sound levels. We differentiate between simultaneous, forward, and backward masking [36], according to whether the probe and suppressor sounds are played concurrently or consecutively. Simultaneous masking is asymmetric in frequency, as lower frequency sounds mask higher ones more likely. Forward masking relates to loud sounds suppressing weak sounds within a delay of tens of milliseconds [86]. Backward masking is merely the opposite of forward masking, not closely as significant, because the encoding of attack is reliable in the auditory system.

**Influence of blindness**   Lack of vision causes developmental differences in both the auditory and visual areas when compared to the cortex of sighted. As the major development period of the brain is confined to younger ages (less than 6 years), the onset of blindness, whether it happens before, called early-blind (EB), or after, called late-blind (LB), influences to a great extent the functional organization of the brain. CB and EB enjoy advantages in auditory discrimination. They are able to recognize rapid pitch and pitch-timbre transitions [87, 88], being 10 times faster in such tasks. On the other hand, the LB lags behind and seems to be inferior in a non-significant manner in pitch change detection compared to both EB and sighted groups [88]. In fact, the earlier the onset of blindness, the better is the pitch discrimination performance [87]. Aligned with these results, we see the tonotopic area of the blind increase substantially, by 84% [89]; though this study neglected to separate measurements from early and LB subjects. EB produces weaker neural responses to pure and frequency modulated simple auditory stimuli [90], as a sign of better processing efficiency and greater ability to disregard irrelevant auditory stimuli in the absence of vision. However, they show more neural activity in complex audio discrimination tasks, when matched with LB and sighted. As we discuss in a later section, cross-modal plasticity allows auditory processing to be relayed to other sensory areas like the visual cortex. Pitch attributes are partially processed in the ventral visual pathway in CB [45].

In summary, the following tonotopic coding features are crucial to consider, in case we plan to feed information to the cortex via the auditory system: 1) the cochlea performs logarithmic compression in the frequency domain, 2) A1 neurons respond to phasic changes more reliably than to sustained stimuli, 3) harmonics of different ratios are encoded separately, but the fundamental frequency and its harmonics are perceived as a single auditory entity, 4) tones of close frequency are likely to be suppressed, 5) multiple tones played at the same time introduce distortion tones, 6) the early and LB show different performance in pitch and timbre discrimination tasks.

### 2.3.4 Binaural spatial localization

As Wenzel [91] described it: "The function of the ears is to point the eyes". Monaural and binaural cues in conjunction aid the approximate localization of sound sources. Once we turn the head towards the source, both our auditory and visual cues become more spatially accurate. Before the sound signal arrives at our ears, it interacts with the floor, torso, shoulders, pinnae and the head. These interactions shape the spectral content of the sound entering each ear, producing monaural cues. Such cues mostly define the elevation of the sound source and whether it originates from the front or the back [14]. On the other hand, binaural cues arise from the integration of sound information arriving at both ears, which conveys information primarily about the azimuth of the sound location.

Lord Rayleigh introduced the duplex theory of binaural localization [92]: ITD dictates in the low-frequency regime whilst ILD dominates at higher frequencies. ITD processing, which can be thought of as interaural cross-correlation, is distinctively performed in MSO [50]. The wavelength of the incoming sound and the distance between the ears determine the efficacy of ITD based azimuth coding. ITD error is the lowest for sounds between 700–1000 Hz, while above 1500 Hz it fails to be informative for humans [93]. Sensitivity to ITD is uniform along the horizontal plane, but as the change in time difference decreases towards the periphery, sound localization accuracy also declines [94]. Apart from the more complex head related transfer function described in the work of Smith and others [94], the Woodworth formula [95] provides a simple frequency-independent definition of ITD as the function of azimuth:

$$\text{ITD}_\theta = \begin{cases} \frac{r}{c}(\theta + \sin\theta), & 0 \le \theta \le \frac{\pi}{2}, \\ \frac{r}{c}(\pi - \theta + \sin\theta), & \frac{\pi}{2} \le \theta \le \pi, \end{cases} \tag{1}$$

where $\theta$ specifies the azimuth, $r$ is the head radius, and $c$ stands for the speed of sound.

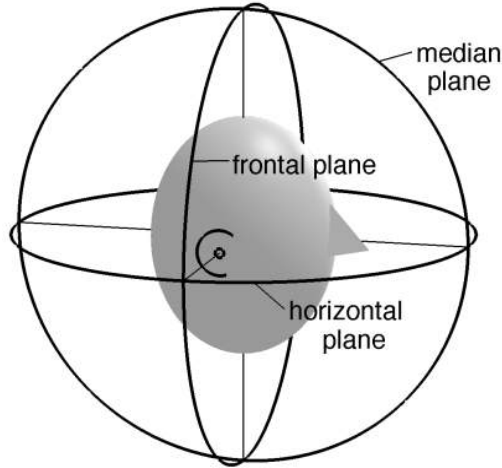

Figure 6: Azimuth and elevation address the interaural polar coordinate system as latitude and longitude mark the surface of Earth. Zero degrees of azimuth is defined as the meridian plane, while zero degrees of elevation effectively refers to the horizontal plane. Figure adapted from the work of Lyon [14].

ILD, processed mainly in LSO, emerges from the head shadow effect: sound is obstructed when passing through the head, which attenuates the overall amplitude and dampens higher frequencies. ILD is mainly evaluated for frequencies higher than 1500 Hz. ITD and ILD cues are already combined in IC, hence AC likewise encodes spatial information, regardless of frequency content [50]; the binaural cues are analyzed as early as $100 \pm 150$ ms after stimulus arrival in AC [96].

Sound localization accuracy varies by azimuth, elevation, spectral content and training. Both azimuth and elevation localization deteriorate towards the periphery. Azimuth coding accuracy is highest at the center and declines with sources of lower or higher elevation [97]. Localization error of azimuth stays below 5° along the midline, between −45° and +45° azimuth; it reaches 15° at 90° on the lateral sides. Azimuth coding in general is more accurate than elevation discrimination, and exploits more corresponding neural resources [96]. In a short noise burst localization task, the spherical error of the two polar coordinates in average was shown to be 5° straight ahead, 10° at 45° azimuth and 12° at 90° [97].

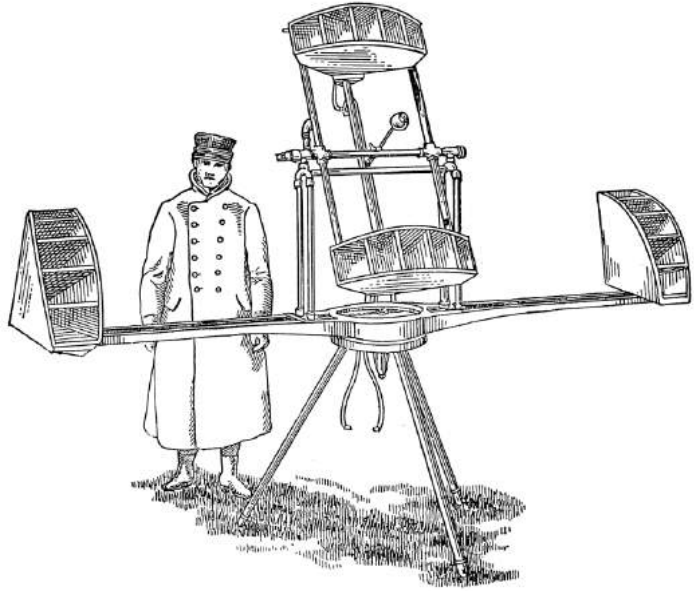

Figure 7: An acoustic goniometer from World War I. It served as a hearing extension, enlarging ITD cues, to gauge distant aeroplanes. Sketch reprinted from the book of Lyon [14].

Smith and colleagues [94] demonstrated the influence of spectral content on sound source localization by playing noise bursts of different bandwidth. Increased bandwidth implies better accuracy until reaching 1–2 octaves. Band noise of $\frac{1}{20}$ of an octave is localized more successfully than pure tones with the same central frequency. Sounds centered on 2000 Hz implies the highest localization error compared to 250 Hz and 4000 Hz. In general, azimuth discrimination accuracy is worse in the range of 1000–3000 Hz.

Identification of binaural cues follows the Haas, or precedence effect: ILD and ITD cues are evaluated for the first 2 ms of the stimulus, and the rest mostly is ignored until 40 ms [14]. Aligned with the Haas effect, longer stimulus does not imply superior localization accuracy [98]. As a final note on the efficacy of binaural spatial coding: it can be improved by training, as shown by the work of Majdak and colleagues [99] in a pointing task.

When compared to the performance of the sighted, the visually impaired localizes sounds more reliably in the peripheral fields [89]. Such higher accuracy may be attributed to the tendency of CB to recruit the right dorsal occipital stream for high-level auditory spatial discrimination [45].

Binaural spatial cues have been widely adopted in V2A SS design, especially the azimuth dimension. If we aim to convey information via sound source location, it is crucial to understand the limitations of and the factors influencing auditory spatial coding, so that the information loss is minimized while the binaural bandwidth is fully utilized.

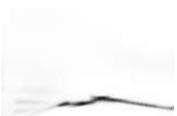

### 2.3.5 Auditory streams

The higher we ascend along the auditory pathway, the temporal window of encoded features widen. Even though the primary AC is organized in a tonotopic manner [23], higher-level stimulus qualities are more accurately represented in its response than the presence of a certain pitch [21]. Neural activity in IC already corresponds to spectro-temporal shifts. Frisina [21] confirmed that by inheriting such features, A1 is more sensitive to abstract auditory entities, such as bird chirps, echo of chirps and ambient noise [22]. Same was shown for music, tonal and speech features [23]. In the following we assess the factors, along which spectro-temporal patterns and auditory entities vary. The gained insights lead us to a more complete exploitation of the A1 coding space, yielding ways to construct higher bandwidth SS stimuli.

**Amplitude and frequency modulation**   The acoustic envelope cues, or amplitude modulation (AM), stand for the change of amplitude across time, while the acoustic carrier cues, or frequency modulation (FM), entail the temporal shifts in frequency. Both of these auditory characteristics are crucial to decode speech and speaker identity [20] and are encoded in AC [24, 25].

Human detection of AM is highly accurate and could reach 1000 Hz, while the discrimination between AM soundscapes of different shape does not surpass the 100-Hz modulation rate [24]. In unanesthetized primates, even 3-ms-long amplitude ramping and damping changes are encoded [100]. AC of humans is further sensitive to as rapid as 64 octave/s of FM [25]. AM and FM encoding schemes demonstrate high efficiency of signal transfer in noisy neural or electromagnetic environments alike. An automatic, absolute perceptual mapping exists between AM rate and visual spatial frequency [26]: a high-density corrugation surface is more likely matched with fast AM rate. Application of this intuitive mapping is beneficial in V2A SS design.

**Stream formation**   A sequence of sound perceived as a singular entity is called an auditory stream. Fusion refers to the merge of a string of sound bits perceived to originate from a single source; fission follows when the audio signals are segregated into separate streams.

The occurrence of fusion or fissure is essentially dependent on the fundamental frequency, phase spectrum, temporal envelope and sound localization qualities of the sound sequence [101]. Moreover, if the listener is instructed to hear multiple streams, they are more likely to do so. Short time interval between consecutive tones of the same pitch motivates fissure. Spectral similarity rather than pitch similarity supports fusion. Abrupt changes in the sound signal resets the stream segregation mechanism and aids subsequent fissure [101].

Ingredients and formation of sound streams are relevant to the design of V2A SS devices, which may aim to recruit the striate cortex to process auditory information. The visually impaired delegates auditory discrimination of complex and attended streams of audio to the visual areas, less abstract features are evaluated earlier in the hearing system [33, 102].

### 2.3.6 Computational models

The compression mechanism of the auditory system is highly complex, as elaborated in the previous sections. Variables such as spectral content, loudness, temporal modulations and location of sound are all interdependent in the accuracy in which they are neurally encoded. Hence, explicitly defining the distribution of these variables, conditioned on the visual information they should convey, harbors extraordinary difficulty. Explicitly defined V2A conversion functions thus tend to produce auditory stimuli that fail to contain entirely the encoded visual information once compressed by the auditory pathway. However, by embedding a computational hearing model, we can implicitly force a learning environment to converge to soundscapes which are resilient to the auditory compression bottleneck.

The learning environments considered in this study are deep neural networks (DNN). A hearing model embedded in a DNN needs to support backpropagation by providing the derivative of its input–output function, so the error gradient can be propagated backwards through the model; i.e. it has to be differentiable. Furthermore, learning in highly recurrent networks tend to be slow and unstable: recurrent structures prevent the parallel evaluation of the gradient and the input-output function, while feedback connections prolong the the chain of gradients to be computed, inciting, for example, the vanishing gradient problem [103]. Therefore if possible, the hearing model applied should not incorporate such highly recurrent architecture.

However, most hearing models consists of infinite impulse response (IIR) filters, which involve feedback connections at different time scales: previous outputs of the filter are fed as input in the subsequent iterations. As a contrast, finite impulse response (FIR) filters can be implemented as causal temporal convolutional networks [104], as they lack the feedback loops mentioned above and they effectively function as a moving average. The following functions illustrate the difference equations of FIR and IIR filters:

$$y_{\text{FIR}}(n) = \sum_{k=0}^{n} b(k) * x(n-k), \tag{2}$$

$$y_{\text{IIR}}(n) = \sum_{k=0}^{n} b(k) * x(n-k) + \sum_{j=0}^{r} a(j) * y(n-j). \tag{3}$$

Soundwave propagation from the environment to the ear is commonly modeled by head-related transfer functions (HRTF), which describe the pressure at both ears given the acoustic wave [14]. HRTFs are predominantly linear and parametrized by sound source location, head shape and ear positions [94]. Three computational hearing stages are commonly distinguished within the ear: outer- and middle-ear, the cochlea and finally the mechanical-to-neural transduction phase [105].

The outer- and middle ear stage is primarily linear and mostly emulated by FIR filters and frequency-dependent gain functions [106]. In the auditory system, the extremely nonlinear mechanisms within the cochlea pose the first substantial modeling difficulties: a complete model should incorporate frequency and amplitude compression, level-dependent frequency selectivity, inter-channel distortions caused by traveling waves on BM, and simultaneous, forward and backward masking [106], among others.

Cochlear mechanics are described by parallel, or cascaded filterbanks, the latter being more biophysically accurate [107]. Saremi and others [108] argue that the cochlear contribution to the human perception is the most influential in auditory processing; an authentic computational cochlear model could simulate substantial portion of the nonlinear aspects of hearing.

IHC perform the next phase of mechanoelectrochemical transduction, which in turn, excites AN fibers. This process results in the quantization of the auditory stimulus, which discards all the information between the quantiles of discrete neural coding [105]. A half-wave rectifier mimics IHC mechanics [14], while AN responses may be further interpreted as a low-pass filter followed by modulation filterbanks [109].

In the following, we enumerate the latest computational hearing models, the biophysical and perceptual phenomena they emulate, the filters they incorporate and their possible applications in learning environments, particularly in DNNs.

**Hearing models**  The most widespread features extracted in computational speech processing tasks are the mel-frequency cepstral coefficients (MFCCs) [14]. Originating from the mel frequency scale of pitch, MFCCs compose a nonlinear spectrum of a sound spectrum logarithmically scaled by frequency. As the representation of pitch in the frequency domain is incomplete, and as MFCCs dispose of phase information entirely, the fine acoustic temporal characteristics are absent from MFCCs [14], which is reflected in our inability to reconstruct the original sound from these features without substantial perceptual degradation.

Gammatone filters were designed after AN fiber responses [110]. By combining Gammatone filters, we may construct filterbanks. Each filter is placed on a CF with a linear, symmetric frequency response. Even though such filters are easy to implement and are computationally cheap, they cannot simulate level-dependent frequency selectivity and the negative skew of AN frequency responses [109]. Gammachirp filters were invented to mend these frequency-nonlinearity and compression deficiencies [111].

The hearing model of Chi and others [112] performs a spectrotemporal analysis of the auditory stimulus by applying an affine wavelet transform that is equivalent to a bank of bandpass filters. This spectrotemporal analysis still models the cochlea as a linear process, hence it misses level dependent frequency tuning and spectral masking features.

Lopez-Poveda and Meddis [106] improved on traditional Gammatone parallel cochlear filterbank solutions by replacing each Gammatone module with a dual resonance nonlinear (DRNL) filter. DRNL emulates some of the nonlinear processes in the BM: simultaneous two-tone suppression is featured, but not for tones distant in their frequency. DRNL adopts cascades of IIR Gammatone and Butterworth filters with a broken-stick nonlinearity.

The computational auditory signal-processing and perception (CASP) hearing model [109] includes a filterbank of DRNL as its BM stage. Additionally, CASP combines a squaring expansion, an adaptation stage realized with feedback loops of different time scales, a modulation phase preceded by a first-order 150-Hz low-pass filter to attain human-like sound envelope sensitivity, and finally, internal Gaussian noise prevents unrealistically accurate acoustic representations. CASP inherits the nonlinear aspects of DRNL.

The cochlear model of Lyon [113], the cascade of asymmetric resonators with fast-acting compression (CARFAC), applies a cascade of second order filters instead of a parallel filterbank, which mirrors a more biophysically accurate propagation of traveling waves along the BM. The cascade of filters realizes a highly recurrent structure, which renders its application in DNNs computationally demanding. Nevertheless, CARFAC delivers masking effects and the level-dependent frequency tuning nonlinearity, and performs exceptionally well in reproducing human cochlear mechanics when compared to other hearing models [108].

Transmission line models are also notable for their biophysically accurate realization. For instance, Verhulst and others [114] emulate both forward and reverse traveling waves in the cochlea, and include detailed IHC, AN and brainstem auditory processing stages. Similarly to cascaded filter approaches, transmission line model architectures are excessively recurrent, and require even more computational resources.

**Applications in machine learning**  Deep learning networks frequently employ hearing models, but solely as a preprocessing step. For instance, Baby and Verhulst [107] filtered the auditory signal through five different peripheral hearing models for comparison, before providing the extracted features to neural networks. As such features describe the perceptual properties of the auditory stimulus more effectively than the raw signal, this approach yields to speech recognition and noise suppression applications.

Hearing models inside learning environments, specifically within the layers of DNNs, could serve two purposes: 1) if their parameters are not optimized, but set to mimic human hearing, they could realize a bottleneck in the network that would implicitly force the audio encoding neural layers to produce humanly perceivable auditory signals; 2) if their parameters are optimized in the learning process, then hearing models could serve as an acoustic feature extraction module. The latter is arguably implemented in the WaveNet neural network [115] at an abstract level; WaveNet could be perceived as a highly nonlinear IIR filter.

## 2.4 Multimodal coding

### 2.4.1 Cross-modal plasticity

Extensive research has been conducted about cross-modal plasticity in the visually impaired [4]. In the absence of visual input, ample evidence exists of AC delegating signal processing responsibilities to the occipital cortex.

**Visual-to-auditory**  Spatial processing of auditory signals, i.e. sound source localization, is partially performed in the right occipital cortex, more specifically, in the dorsal and lateral ventral parts [39, 116], for CB. The inferior parietal lobule seems to mitigate the auditory spatial information for more demanding computation towards the occipital areas [116]. Similarly, Collignon and colleagues [45] reported activations of the right cuneus and the right occipital gyrus in CB. These reports suggest that visual experience is not necessary to develop the spatial processing nature of the visual dorsal stream, as such networks are involved in visual spatial computation in the sighted, as well. Experiments done with V2A SS devices showed similar results [42]. Even after 1.5 hours of SS training, the visual dorsal stream of CB responded most to location information encoded in the audio [117]. Depth processing of SS auditory stimulus reported to be performed also in the areas of occipital cortex responsible for visual depth perception [118]. Furthermore, the blind partially delegates auditory spatial motion processing to visual areas with the same motion processing responsibility originally documented in the sighted [119].

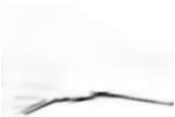

SS research further yielded results in demonstrating how shape information is processed when presented as auditory stimulus. fMRI measurements were conducted on sighted, LB and CB. Results demonstrated that the lateral-occipital tactile-visual area (LOtv) responded to shape information in all groups, but only for those subjects who were trained to interpret the SS auditory signals [41]; LOtv originally reacts to visual shape information in the sighted. The same study reported activation in the posterior occipital cortex analogous to retinotopic visual regions, while performing the same shape detection task. Striem-Amit and others [117] also confirmed widespread activations in the visual ventral stream of CB, including LOtv, during a shape discrimination experiment.

SS devices designed for the visually deprived, e.g. TV [9] and BrainPort [120] induce activity in the visual cortex when used after training [40] in EB and CB. Specifically, vast research has shown activations in the lateral occipital cortex (LOC) during V2A SS tasks [41,121,122]. LOC activity may not be enhanced by SS training, but its functional connectivity to auditory areas may increase post training. Merabet and colleagues [44] demonstrated in a case study that repetitive transcranial magnetic stimulation (rTMS) delivered to the striate cortex impairs the TV-trained blind in a visual task, indicating the recruitment of occipital areas for auditory signal processing. After only 2 hours of V2A SS training, Striem-Amit and colleagues [33] reported that the blind delegated word processing to the visual word form area when reading. Hence, when visual information of text is presented as audio signals, the blind's brain rapidly rewires itself to process the demanding stimulus. Similarly, Kujala and others [102] postulated that the primary visual cortex is only involved in the processing of higher level auditory streams that are specifically attended to. In their experiments, Finnish vowels and vowel-like tones were played, standard and deviant ones in succession, the latter needed to be detected by the subjects. Here, the occipital areas responded only to cases, when the standard and deviant stimuli differed slightly, regardless of being vowels or tones. This demonstrates the tendency to delegate only difficult acoustic signal discrimination to visual areas that cannot be solved solely by AC. On another account, reception of vocal and non-vocal stimuli was examined in sighted, LB and CB [123]. Both blind groups showed visual cortex activations, regardless of stimulus type. Furthermore, the fusiform face area of CB activated only in response to vocal sounds, which supports the intuitive idea of voices being the "auditory faces" for the blind.

**Visual-to-tactile**   Research done in visual-to-tactile cross-modal plasticity further attests the recruitment of the visual areas by other sensory systems. The occipital cortex of the blind is shown to respond to Braille reading [124–127]. Merabet and others [125] extended the research by including blindfolded subjects in a Braille reading experiment, and showed improvement on tactile task performance after 5 days of visual deprivation. Moreover, rTMS disruptions fired at the occipital cortex worsened the performance only for the blindfolded case, but not at baseline, nor after the removal of the blindfold. We need to make a difference between worsened performance and disrupted reading, as the latter has been only induced in CB by TMS [128], but not in LB. In a tactile discrimination task of bottles and shoes [129], neural activity was found in the inferior temporal cortex, higher up the ventral visual pathway, even for sighted participants without visual deprivation. This suggests that the absence of visual input is not a necessary precondition of cross-modal plasticity of higher visual areas. Another Braille reading study [124] demonstrated that the striate cortex is only acquired to process attended higher and more complex tactile information; it was not activated by simple finger tapping or electric stimulation of fingers presumed to mimic Braille reading. Yet in another experiment [130], subjects were trained to use a tongue display for 2 days, while TMS was applied on the visual cortex pre- and post training. Some blind subjects (3 EB, 1 LB) felt tactile sensations on the tongue only post training, demonstrating the cross-modal connections spawned by the SS device.

The deference of computation to sensory areas absent from input is revealed in the deaf, as well [131,132]; primary and higher auditory areas of deaf tend to respond to visual information.

In summary, spatial and shape information encoded in auditory stimulus seems to translate well to corresponding visual areas in the blind, which supports the design of V2A SS devices conveying these very visual features. Color and shade are not among such properties, as both of them are highly dependent on ambient light, which cannot be supplied to the blind via other, i.e. haptic, modalities. It is vital to receive tactile sensory feedback of the translated visual features in parallel to the V2A SS auditory signal, in order to initiate multisensory perceptual learning [13]. Otherwise, in lack of feedback, the visually impaired cannot grasp the hidden information from the substituting stimulus, which drives the brain to generalize over the hidden variable. For instance, shading of a tree in the morning is different from the shading in the afternoon, as the sunlight arrives from another angle. The issue is more pressing indoors, where distinct combination of light sources cast shadows in distinct ways, but the shape of objects remain the same. The complex interplay of reflection and refraction of light further complicates the substituting signal, without any way of integrating these features. On a final note, the striate cortex of the blind is acquired to process discrimination of auditory streams with nuanced differences, while obvious auditory classification problems remains to be solved in the auditory pathway. Therefore, if we aim to relay V2A SS stimulus to the visual areas, we may be encouraged to sample our sounds from a well-defined distribution of soundscapes,

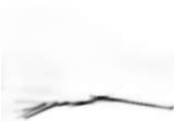

where a nuanced difference in the sound shall correspond to a nuanced difference in the corresponding visual information; all this in the hope of inducing occipital cortex activity and visual perception if possible.

### 2.4.2 Onset of blindness

From the early periods of postnatal development, until about the age of 6 [127], the human brain undergoes rapid development. In this span of time, the brain is extremely sensitive to sensory input, and structure itself accordingly. After this early period, the loss of sight cannot be met with substantial structural changes that would allow it to, for example, rewire V1 to efficiently operate on non-visual sensory signal [39]. Therefore, it is imperative to group the blind corresponding to the onset of blindness, and review cross-modal plasticity studies in the light of such grouping. We call those who are visually deprived since birth, congenitally blind (CB), those who lost their sight before the age of 6, early blind (EB), and those who lost it after, late blind (LB).

As described in the previous section, auditory spatial processing, or source localization tend to be delegated cross-modally to the occipital cortex, or more specifically to the right visual dorsal stream [39], or in the case of another study [133], to the left visual dorsal stream; the reason behind the hemisphere switch between these studies is unknown. In fact, the amount of neural activity in the visual dorsal stream is positively correlated to the performance of sound localization, but only for CB and EB. Although LB does not benefit from the recruitment of the dorsal extrastriate cortex, all blind tend to perform better in spatial auditory tasks like minimum audible angle discrimination [134] and peripheral auditory attention [135], when compared to sighted. In general, CB and EB exceed LB in localization tasks [39]; the age of blindness onset is inversely correlated to the level of performance [133]. If sight is lost early in life, the dorsal visual pathway is recruited to execute the computation of non-visual spatial cues [116]. These results demonstrate that functional selectivity of spatial processing in the visual dorsal pathway is at least not attained in a rapid fashion post visual deprivation [39].

In the blind's V1, functional selectivity to shape processing is akin to the auditory spatial processing case described above. Tactile shape discrimination induced neural activity in V1, if the onset of blindness is before the age of 16; otherwise the activity is suppressed [136]. However, in an experiment where blind subjects had to read nouns through Braille and come up with a fitting verb to them, several occipital sights bilaterally responded both in EB and LB [127].

Along the lines of cross-modal functional recruitment of the occipital cortex, gray matter concentration in EB's primary visual cortex is higher than in LB's [137]. More specifically, there is an inverse correlation between the amount of gray matter and the onset of blindness. This phenomenon can be accounted to the inability of the visual system to rewire itself after the early critical period, which leads to the atrophy of unused occipital areas in LB. On the other hand, reorganization of striate regions in CB and EB may follow through and such areas may specialize in processing information channeled from other modalities. CB suffers from volumetric reductions in the visual pathway [138], as well as in the thalamus and the LGN [139].

Cross-modal connections between the auditory pathway and V1 have been examined in the blind. Functional neural connections between these two modalities are either thalamo–cortical or cortico–cortical in majority, and the latter case is more supported [47]. Within the category of cortico–cortical connections, we can differentiate between direct feedforward or indirect feedback connections. In case of direct connections, the auditory signal is channeled to V1 straight from A1, without any intervention from higher-level networks, while an indirect path would first journey through associative, parietal regions, before arriving at V1 as top-down feedback signals. V2A cross-modal connections are more likely to be direct in CB and indirect in LB; the chance of having both feedforward and feedback connections at the same time is lower in both CB and LB [39]. More specifically, the cross-modal route likely travels across the intraparietal sulcus before reaching V1 in LB. When designing V2A SS devices, establishing the nature of the V2A cross-modal route is vital, because the encoding of auditory features are different in the thalamic and cortical level.

Counterintuitively, studies of EB revealed decreased functional connections between the striate cortex and ipsi- and contralateral temporal regions responsible for auditory processing [140]. On the other hand, functional correlations between the occipital cortex and higher level cortical areas of cognitive control is increased for EB, which has been explained by the recruitment of the visual cortex to aid cognitive functions, like memory, attention and cognitive control. Bock and Fine [140] argue that if such recruitment occurs, we would expect the corresponding connected cognitive and occipital areas to have similar functional roles, which is not the case. The decreased anatomical and functional cross-modal connections and the elevated involvement of the visual cortex with cognitive areas may be alternatively explained by the mixture of experts (ME) architecture. The ME architecture involves the interaction of multiple expert networks and a single gating network. Expert networks, like the visual and auditory cortices, learn to perform certain tasks and compete to perform others. The gating network mediates the competition and divides tasks between the expert networks to achieve maximal efficacy and minimal correlation between the experts; it does so, by inhibiting or disinhibiting the output of the experts given a certain task. Hence, if the occipital cortex of the blind learns a more competent decoding of Braille signs, the gating network assigns that task to the visual network and drives the somatosensory cortex to shift its responses away from that particular task. Therefore, the ME architecture predicts the lower cross-modal and higher visual-to-cognitive connections demonstrated in the visually impaired [140].

Complete absence of visual experience drives the preference for an egocentric reference frame, as non-visual modalities constrain themselves to sequential perception [141]. To attain the alternative, allocentric reference frame, one needs to observe multiple objects at once, sensing them spatially relative to each other. In other words, an allocentric view allows the integration of stimuli, regardless of modality, appearing within the same expanded spatial window. CB lacks such a view and it is arguable whether they could ever attain it.

In conclusion, studies of cross-modal plasticity paints a mixed picture of the potential of V2A SS in visual rehabilitation [142]. On one hand, EB can transfer spatial and shape information to the occipital cortex via straight, feedforward connections, so functionally selective areas may process auditory features in functionally corresponding visual regions; i.e. shape is computed in ventral and sound localization within the dorsal occipital pathway. Moreover, LB may have the retinotopic neural architecture in place to compute audio encoded visual information in parallel, and thus, partially regaining perceptual vision is plausible [10]. On the other hand, without enough early visual experience, developed retinotopy and the ability to form an allocentric view, EB has low, if no, chance to gain synaesthetic visual perception; though the case study of Hofstetter and others [143] supports the possibility to build a form of retinotopy. Furthermore, as LB has only indirect cross-modal connections, a difficulty to recruit the striate cortex and potential atrophy of relevant retinotopic areas, the functional revival of their occipital cortex is questionable; though the recovery of two blind individuals with slight light perception has been attained [10]. Hence, the only group that may achieve the Holy Grail of V2A SS, the synaesthetic visual experience, is LB, with some remaining perception of light.

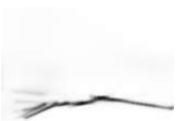

### 2.4.3 Nature of cross-modal connections

As outlined in the previous section, the auditory to visual cross-modal neural highway is cortico–cortical in majority, CB and EB having direct, feedforward connections, while indirect, feedback connections reside in LB. The occipital cortex receives information from the primary AC, and thus likely performs higher-order computation on soundstreams, relative to simple sound discrimination already computed in the auditory pathway [33, 102]. The visual areas are exclusively recruited in a cross-modal fashion, if the acoustic stimulus is attended to, and the novelty detection or discrimination task executed is difficult, nuanced enough to require further processing after A1. Similar mechanism has been shown in the case of somatosensory and occipital cross-modal engagement [124]. The striate cortex does not respond to Braille reading-like sensory stimulation, it participates only in higher-level computation of attended Braille comprehension.

Cross-modal functional connections are likely spawned on two timescales by two different mechanisms: 1) the rapid unmasking of already existing cross-modal neural routes, and 2) the slow reorganization of cortical synaptic network. Fast neuroplastic changes observed in blindfolded studies indicates that the visual cortex can be recruited within 5 days of visual deprivation by other modalities [118,125]. In such a short time, plasticity may only occur via the decline in inhibition of already existing cross-modal connections. Further support for fast unmasking was presented in the study of Nau, Murphy and Chan [40]. Blind subjects only needed 10 minutes of training with a V2A SS device to demonstrate BOLD activations in the striate cortex, while using the device. In another experiment [122], after two hours of training, ERP differences in the posterior occipital cortex were caught at 150–210 ms and 420–480 ms after the onset of the soundscape, when compared to untrained subjects. This result shows how fast the unmasked cross-modal network is able to transfer auditory information to the striate cortex. The gradual reorganization of the synaptic network is expected to follow after the cross-modal masking is disinhibited [29].

Anatomical studies of macaque sensory cortices have shown direct, monosynaptic connections from A1 to V1. However, such neural routes are sparse, having significantly less bandwidth than the feedback connections established with higher, multisensory areas [144].

All in all, the above referenced research points out the cortico–cortical connections between A1 and V1 and the higher-order nature of auditory soundstream analysis the primary visual cortex performs in the context of V2A SS. In order to effectively recruit the striate cortex to process V2A SS stimulus, the presented soundscape should hold all the visual information when encoded in A1. Hence, it may be beneficial to translate visual information into frequency and amplitude modulated audio streams, which are more reliably coded for in the cortical level of the auditory pathway than single sinusoidal signals [20]. Most likely, there exist spatial, temporal and associative relations between the modalities, which constrain the array of potential cross-modal connections [144]. By associating V2A sound stimuli to spatially and temporally equivalent visual features, and further relying on behavioral studies of visual–auditory congruence, we can derive a V2A conversion logic that exploits already existing cross-modal connections more effectively.

## 2.5 Sensory substitution

SS entails the delivery of sensory information of an impaired modality via another, spared sense [145]. In case of V2A SS, sight is substituted by hearing, so visual features, such as spatial position, luminance and chroma, are encoded in audio signals. As reviewed in previous sections, the received sound first climbs the auditory pathway, before being encoded in the primary AC. Some low-level properties of the sound signal are extracted along the way before the occipital cortex is recruited, either directly or indirectly, to further process higher-order characteristics of soundstreams in the blind. Visual areas perform computation on the auditory features encoded in A1. The acquired additional processing power aids the blind's higher performance in sound localization [134], pitch discrimination [87] and perception of temporal changes [88].

### 2.5.1 Conversion methodology

The core of a SS system is the conversion method that mediates between the representation of the substituted and substituting sensory modalities, or the function that translate images to soundscapes in case of V2A SS. This study differentiates between explicit and implicit conversion methods. The explicit approach has an absolute grip on the input space, clearly defining the audio signal for every possible visual feature, usually superimposing the signals linearly across features (Figure 8). For instance, each pixel on an image is converted to a soundscape specified solely by the position and color content of the pixel, then these signals are overlapped [9]. On the other hand, an implicit scheme acts as a black box, employing an intricate, nonlinear, interaction between the visual features present (Figure 9). Meaning, the attribution of a single pixel to the produced sound is neither defined, nor tractable without knowing the state of all other pixels. Computer vision algorithms applied in SS typically yield implicit models, for example, algorithms translating scenes to corresponding verbal descriptions [146, 147].

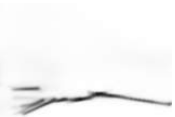

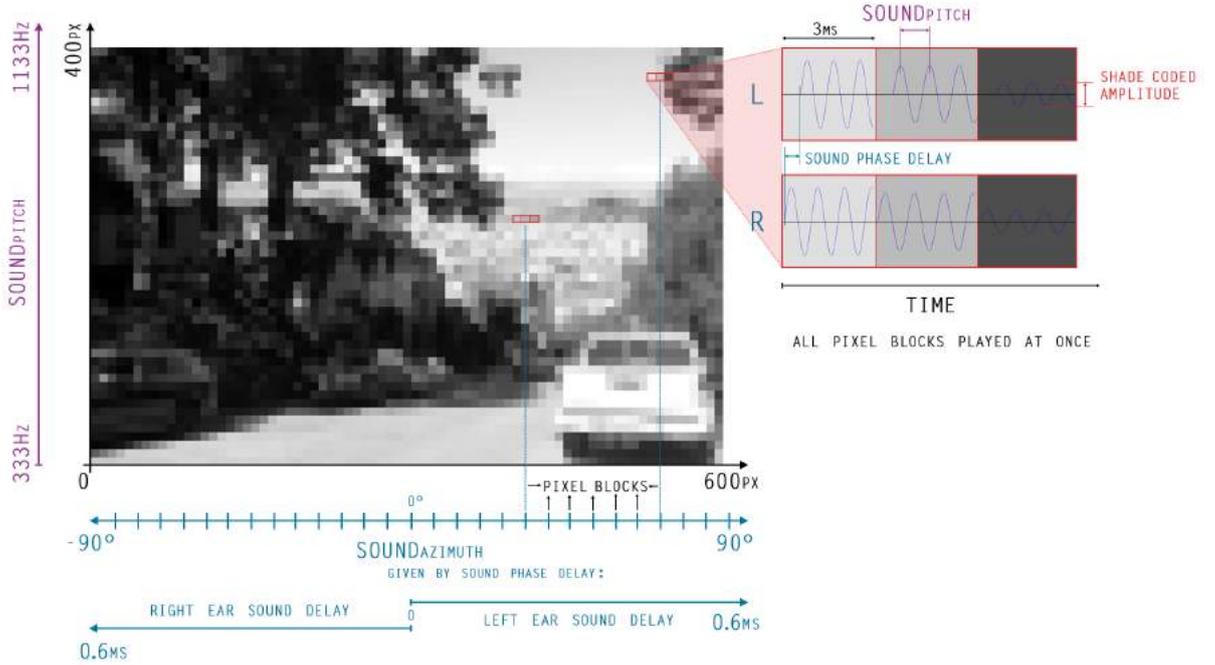

Figure 8: A disentangled example of an explicit SS conversion function, which we preliminary experimented with in the beginning of the study. It plays a certain amount of pixels, here 3, in a sequence, assigning sound sequences to each pixel as TV [9] SS device does. Soundscapes corresponding to all pixel blocks of three are played at the same time, which yields a short, yet rather convoluted audio representations.

**Substitution delay**   Substitution delay spans from the time the substituted information is captured, until the substituting signal is presented entirely. In case of V2A SS, substitution delay begins at the instant an image is taken and ends right after the corresponding soundscape finished playing. Substitution delay includes the computational time of the conversion logic and the duration of the substituting signal. One may further expand the definition by the time the substituting signal is neurally encoded in the human brain, including the period of cross-modal processing. Yet, we have limited knowledge about the timescale of cross-modal processing [122], and the ways in which it could be informative in practice, hence, we treat it as a constant.

**Visual space compression**   Vision encompasses high, parallel spatial and relatively low temporal resolution, while auditory coding is highly sensitive to temporal transitions and predominantly serial in nature. The optic nerve consists of 1 million fibers, whereas AN incorporates only 30,000 [43]. Thus, substituting sight by hearing, encoding high-dimensional visual information into sound, pose as a substantial chal-lenge; total vision restoration by SS is unlikely. The major approach of solving the challenge is visual space compression: we need to discard detailed spatial features, and decrease the temporal resolution by extending the encoding of a single image across time [9]. Moreover, chroma and shading may be removed for the following reasons: 1) such concepts are unintelligible for EB and CB [142], 2) they cannot be integrated with tactile sensory feedback [13], and 3) they would retain relatively high bandwidth [148]. The remaining edges, or contour of an image may not be treated pixel-by-pixel, but may be encoded as a whole, in a similar manner as the primary visual cortex processes bars: by spatial position and angle. This very last step entails the abstraction of pixels to bars of different orientation, which results in the reduction of required coding space, as a single bar represents multiple of pixels; albeit, abstraction occurs on the expense of detail. At the extreme end of this abstraction lies the encoding of a visual object, or even the entire scene, as a single auditory event: for instance, the verbal description of a scenery.

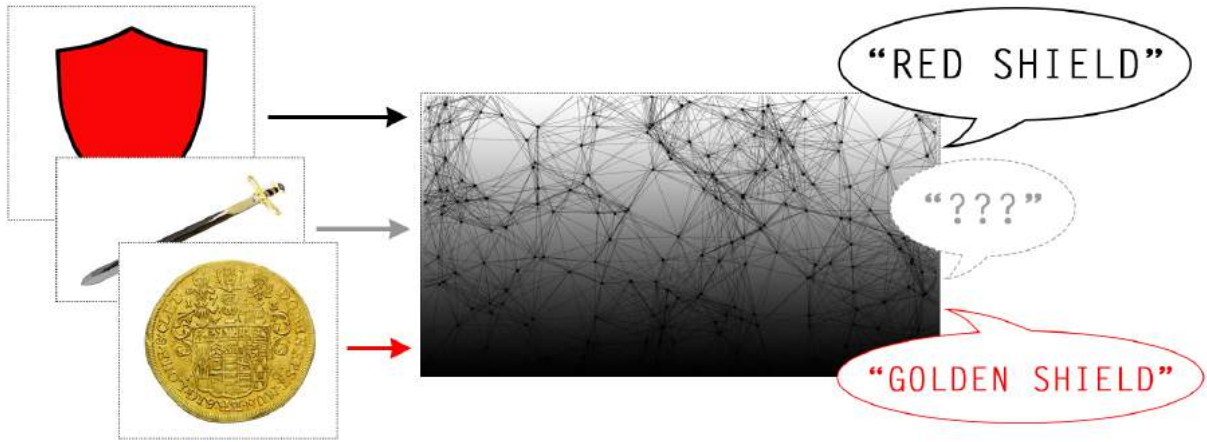

Figure 9: Illustration of an implicit conversion method, trained on the classification of shields, translating the input image to a verbal description, similarly to DEEP-SEE [147]. The conversion logic implemented as a DNN, behaves like a black box: no explicitly defined function exists that could describe the contribution of each pixel to the produced audio, independent of the rest of the visual space. Implicit methods are inclined to fail at the conversion of visual features that it has not been trained on: the hypothetical model baffles at a sword and misinterprets a coin.

In deep learning models, particularly in convolutional neural networks (CNN) [34], we wield methods to assess the response features of filters [149]. As we climb the hierarchy of neural layers, the abstraction of image characteristics increase, akin to the human visual pathway. However, abstraction through code-space reduction is only possible if an underlying pattern exists in the images the CNN is trained on, which is true for natural images. DNNs internalize the pattern presented to them, so they reserve code-space for images they have been trained on, while neglecting the representation of unseen visual features. Deep learning models applied in SS are part of the implicit group, because the conversion function acts as a black box, instead of being clearly predefined.

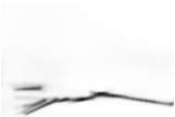

By expanding on deep learning, we identified the superior visual space compression efficacy of implicit methods compared to explicit strategies. To reiterate, this originates from the fact that implicit approaches optimize for coding space on all levels of visual abstraction, while explicit solutions have to stick with a predetermined conversion function that hardly incorporates the distribution of likely visual elements. This inherent advantage of implicit methods is especially critical in V2A SS as the substituting modality holds orders of magnitude lower bandwidth than the substituted one, and thus, every advance in the compression of visual coding space is met with higher auditory coding efficiency. In other words, the further we compress the visual space, the more visual information can be condensed into an auditory signal of a given length; that is, the shorter the substitution delay may become.

However, the comparison between explicit and implicit conversion strategies is more nuanced. Because the logic of explicit methods can be verbally described, they are easier to grasp, which facilitates SS training [32, 33]. Explicit solutions also reliably convert unseen stimulus, while the response of implicit conversion is unpredictable for untrained input. In terms of substitution delay, implicit methods may yield shorter soundscapes due to more efficient visual space compression, but the computational requirement of the conversion logic may be higher, resulting in a longer computational delay. Finally, we believe that the development of explicit approaches has been stagnating in the previous 26 years, regurgitating the same conversion logic in slightly different forms. Hand-crafting SS models is immensely difficult, particularly if we intend to include the ever expanding research in cross-modal plasticity, psychoacoustic and perceptual learning. The barrier of entry has grown incredibly high for explicit methods, while implicit solutions, in combination with deep learning, show great promises with regards to the available tools and technology. In other words, an implicit conversion function is optimized according to the boundaries we impose; we barely need to care about the specificities of the model structure and the internal transformations, if we can define the loss function appropriately.

Table 1: Explicit versus implicit SS conversion methods.

|                          | Explicit      | Implicit         |
| ------------------------ | ------------- | ---------------- |
| Visual features          | any feature   | trained features |
| Visual space compression | predetermined | data specific    |
| Conversion logic         | transparent   | mostly black box |
| Verbal description       | expressible   | semi-expressible |
| Barrier to entry         | formidable    | low              |
| Computational demand     | low           | moderate         |

In the following sections, we review the SS devices designed so far, categorize them according to the level of visual space abstraction they employ, contrasting implicit and explicit conversion methods. Furthermore, we explore SS training procedures in detail, before concluding with the achievements and limitations of V2A SS devices.

### 2.5.2 Sensory substitution devices

In this section, we review V2A SS devices. We deliberately pay exclusive attention to V2A conversion methods, omitting tactile to visual SS [145] and approaches solely conveying depth perception, for instance, by echo location [150]. For each SS device, we describe the following three major components: 1) the preprocessing function of visual features, applied to compress visual space, 2) the V2A conversion logic, and 3) the audio synthesizer.

One of the most widely used SS visual prosthesis is The vOICe (TV) system [9]. Although, the device was designed in 1992, its major characteristics are mirrored in even the most recent V2A SS solutions [146, 151]. Images are first downsampled to a 64–by–64 grayscale representation (later upscaled to 176x64), containing 16 gray tones. Each pixel is explicitly converted to a sinusoidal signal, which is defined by the spatial position of the pixel and its luminance. The amplitude of the sinusoidal represents the gray-tone content. Vertical position is translated to an exponential frequency distribution, reflecting the spectral coding of the cochlea. Soundscapes of a single pixel column are played simultaneously, while the whole image is scanned from left to right across time, such that the horizontal position is denoted by the soundscape delay. Panning of the audio signal is further introduced to assist horizontal discrimination; pixels on the left are perceived stronger in the left and pixels on the right are heard more robustly in the right auditory field. A distinct cue sound denotes the beginning of every image translation.

The next major contestant in V2A SS, here named Real-Time Enhancement of Vision Rehabilitation Using Auditory Substitution (RTEVRUAS), applies a different image preprocessing and conversion method than TV [152]. Visual space is downgraded to 124 grayscale pixels, with an 8–by–8 pixel sized fovea region and a 60 pixel periphery, depicting the high central and sparse peripheral receptive field distribution of vision. Similarly to TV, loudness implies brightness, pitch codes for the vertical position, and stereo panning conveys horizontal location. However, soundscapes corresponding to all pixels are superimposed at once without delay. Playing the auditory stimulus of a whole image at once is made possible by the association of distinct frequencies to every pixel. Moreover, pitches are allocated so vertically neighboring pixels respond to harmonics, while auditory representation of rows of pixels correspond to non-harmonics. A clever approach to exploit the encoding of harmonics in A1 [75].

The first attempt at including color information in the V2A SS repertoire originates from the design of See ColOr [6]. This device quantizes hue into 7 colors, corresponding to 7 specific musical instruments in the auditory domain. Bologna and colleagues reasoned for the reliance on such a transition function by analogy, arguing that "colour of music lives in the timbre of performing instruments". They suggested that hue information serve to merge visual features of mono-colored objects to be detected as one coherent entity, while also providing practical, additional knowledge about whether the user stands on green grass or not. In the preprocessing step, image segmentation is deployed, which simplifies, blurs the visual space, but does not compress it, as neither the size, nor the color depth of the image is reduced. Although, segmentation indeed lessens variability by removing detail, such details are likely to be imperative to cross-modal learning: e.g. edges of objects if obscured, the amount of handy tactile feedback shrinks, which is necessary to associate object shape and the corresponding soundscape. Visual space reduction is achieved by only sonifying a small subwindow in the center of the image (25 pixels out of a 320–by–240 image), while a saliency detection algorithm is run on the rest of the visual field. In order to drive the user's attention towards unique visual features outside of the sonified subwindow, distinct spatialized alarm sounds denote the location of such salient objects. See ColOr introduced sound elevation into the V2A coding scheme, applying HRTFs. Horizontal position is encoded in the combination of ITD and ILD cues, while vertical pixel location is mirrored in sound elevation.

The Vibe [8] was developed on the premise of a V2A SS framework, in which several different conversion functions can be tested, primarily for research purposes. In its essence, Vibe translates RGB images into soundscapes. Receptive fields are defined, each responding to a group of pixels, via an arbitrary function (e.g. mean of pixel values). By implementing ITD and ILD cues, a binaural soundscape is derived from the sum of receptive field states, playing the whole image simultaneously.

Since the Vibe, explicit conversion method design have barely improved. Eye-Music [151] combines See ColOr and TV, performing color clustering, synthesizing soundscapes column-by-column, and ceiling the frequency at 1568 Hz, assuring the pleasantness of the auditory stimulus; albeit, the frequency cut-off reduces the auditory coding space by half.

In the reign of implicit conversion functions, research endeavors have been scarce. See ColOr [146] relied on a DNN to recognize and classify characters, before verbally pronouncing them. DEEP-SEE [147] builds on more sophisticated computer vision algorithms to track, and deep CNNs to detect objects, including cars, bicycles, and obstacles. Discovery of an object is followed by a verbal description, delivered through a bone-conducting headphone. DEEP-SEE falls at the extreme end of visual space reduction, as the images are compressed down solely to a set of detectable objects and their positions.

As demonstrated, explicit conversion methods dominate the V2A SS field, with just a couple of implicit transition functions around. Among explicit strategies, pitch denotes vertical position of visual features, while time delay and/or binaural cues represent horizontal location. Translating luminance to signal amplitude is common, and colors are consistently encoded in timbre. Visual space compression is performed by pixel downsampling, computer vision algorithms, DNNs, and color clustering. There seem to remain an uncharted territory of implicit conversion techniques, with a low-to-medium visual space reduction function.

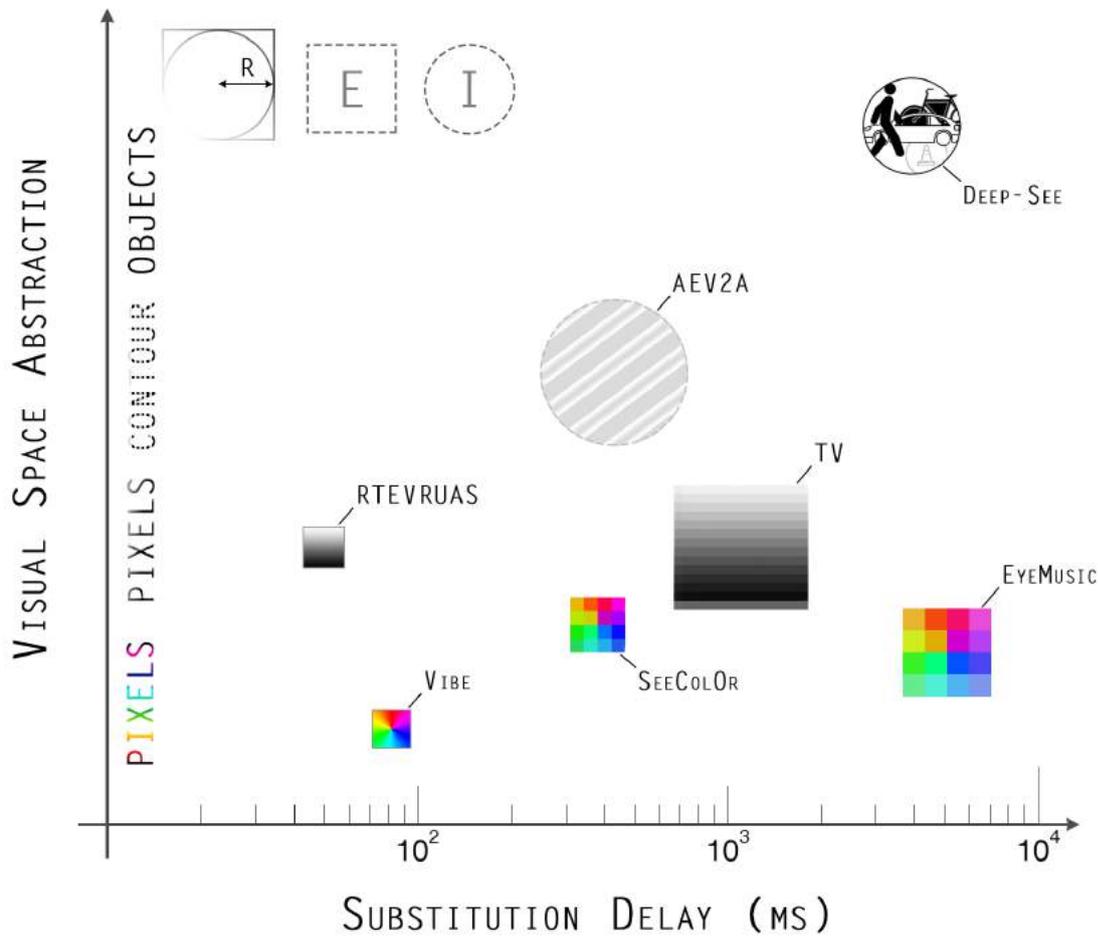

Figure 10: SS devices spread along the axes of substitution delay and visual space abstraction. Substitution delay covers the duration of time from the point an image is taken, until the corresponding soundscape is played in its entirety, including the computational time of the conversion method. Visual space abstraction loosely delineates the amount of visual detail removed from the images before translated into audio. Square symbols represent explicit, circles depict implicit conversion approaches. The size of these symbols, $R$, displays the amount of perceivable, conveyed visual space, i.e. the dimensions of the converted images. The Vibe and RTEVRUAS SS devices superimpose the sound sequences of every pixel simultaneously, which leads to convoluted soundscapes and thus to limited perceivable visual information. Visual space abstraction spans through colored pixels, clustered coloring, grayscale, contour lines, up to the representation of single objects or scenes. Our proposed conversion logic, AEV2A, extracts the contour information of images before transforming them into soundstreams. Ideally, we would aim to achieve as low substitution delay and as much conveyed visual space as possible. In V2A SS, a compromise need to be made between these two factors, which this figure attempts to depict.

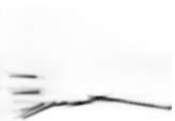

### 2.5.3 Developments

Apart from the variety of V2A SS devices introduced since 1992, numerous incremental improvements to such systems have been proposed in terms of image processing, audio sonification and cross-modal stimulus conversion.

Developments in the preprocessing of SS image inputs tend to draw analogies from the visual system of the sighted. RTEVRUAS [152] artificially simulate the receptor distribution of the retina, while the design of Buchs and others [7] lets the user manually zoom-in the center of the visual field. Kishino and colleagues [148] ran an edge detection algorithm on the input image and translated solely the remaining contour to sound. Kishino and colleagues established that the retina performs a preliminary step to edge detection, hence validating the computational execution of this step for those without a functioning retina. The choice of contour-only conversion also builds on the previously presented cross-modal plasticity research, indicating that auditory information processed in the visual pathway arrives at V1 [47]. Another advantage of visual space abstraction at this level is the removal of details, like shade and color, that are difficult to integrate with tactile information, impeding the cross-modal learning process [13]. Further instances of SS image preprocessing techniques include the application of clustering [6] and machine learning algorithms [147], primarily aiming at visual space reduction. Finally, by benefiting from depth cameras, we may filter out visual features beyond a specific range to adjust for the peripersonal-dominant perception of the visually impaired [142].

Regarding advances in sound generation, the delivery of binaural cues has enjoyed exclusive attention. Simple audio panning was improved by the debut of HRTFs, which provide means to simulate ITD and ILD cues in a frequency-dependent manner, so spatial elevation of soundscapes can also be discriminated between. Bujacz and others [153] found that personalized HRTFs yielded no substantial advantage over generic ones in SS.

In order to optimize the V2A conversion method of TV, Wright and Ward [154] ran a genetic algorithm tied to the performance of sighted participants, who were solving SS tasks. The parameters of TV were genetically evolved as the tasks got evaluated. Although, they only assembled a modest collection of samples, they proposed that elevation with pitch and luminance with loudness are congruent, so high pitch should correspond to visual features at high elevation, and high luminance to loud soundscapes. However, this study neglected including the correspondence of loudness and object size [19], which might have overruled the previously mentioned association to brightness.

In terms of the advancements in the user interface design of SS devices, we see the miniaturization of electronics and improved battery power to be the major drivers [155].

### 2.5.4 Training

Training procedure is critical in the success of V2A SS devices escaping out of the lab environment [10, 142]. A steep learning curve and a lack of gamification elements may cause users to withdraw from further practice [142]. SS training guides cross-modal plasticity both in the fast unmasking and the long-term establishment of new neural connections. By selecting from the myriad of available procedures, we can accelerate the initiation of cross-modal connections, saving time and effort for the blind. Training is also closely tied to the efficacy assessment of SS devices, as the specificity of the learning and the assessment method imply the user's ability to generalize SS in real-world environments.

Although the ability to classify abstract shapes and complex objects or to navigate have been demonstrated with SS devices [27], the generalization of the learned skills still constitutes the core problem. This is reflected in the reverse hierarchy theory, which suggests that the difficulty and specificity of a SS training procedure govern the level of cortical processing and the ability to generalize [13]. In V2A SS, the more processing-heavy the discrimination between soundscapes is, the more the computation is driven cross-modally to the primary visual areas, instead of being resolved higher up in the visual pathway. Moreover, generalization is more successful for learned associations processed in higher perceptual areas, while the lower-level regions tend to specialize in the stimuli presented during training. Hence, the difficulty of SS training needs to scale with the variability of the introduced stimuli: e.g. in a face discrimination task, the number of faces presented should far exceed the number of triangles showed in a simple shape classification exercise. As further discussed in the work of Proulx and others [13], humans can generalize to untrained auditory frequencies, but not to new temporal intervals. The former encourages the encoding of visual features into sound frequency of high variation, while the latter hints at constraining the length of soundscapes; by keeping it constant, any image is translated to a constant length auditory signal.

The field of multisensory perceptual learning resolves the ways in which perceptual plasticity promotes improved representation of the sensory environment [156]. A well-documented catalyst for perceptual learning is sensory feedback as a form of reinforcement [157]. In case of attaining visual perception through auditory information, haptic feedback substantially facilitates the learning process [13], which encourages the design of multisensory training protocols. Furthermore, Elli, Benetti and Collignon [155] suggested that SS stimulus should only contain perceptual information that has been learned before through a spared sense. For these reasons, SS training in a virtual environment should be avoided [13], though Xu and others [158] presented promising results. As the technology backing virtual reality (VR) is maturing, the application of VR in SS vision rehabilitation protocols becomes increasingly appealing. Building VR spaces offers an inherent cost and scaling benefit [150, 159], compared to the design of real-life settings [160, 161]. However, as such virtual environments lack tactile feedback, V2A training needs to resort solely to simplistic shapes and surroundings, such as walls and cylinder obstacles [150]. Anything more complex would require haptic presence or ample verbal description, with the latter potentially proving to be unreliable.

The numerous solutions providing haptic feedback can be arranged into two groups: mechanical and electrical. Mechanical approaches include two-dimensional and three-dimensional shapes, possibly 3D printed, with a high contrast, as with white objects placed on a black magnet board [3]. Stiles and Shimojo [162] adhered cardstocks to a desk surface to represent patterns, leveling the surface, so the black background and the white foreground of SS stimuli are haptically discernible. More advanced methods of material science may construct surfaces of unequivocal precision [163] to represent tactile stimuli of high fidelity. Electrical tactile displays can employ complex-shaped, two-dimensional, haptic feedback, but they cannot represent depth. Senseg [164] and Disney's TeslaTouch [158] create two-dimensional tactile images that the visually impaired could associate with the corresponding soundscapes in a V2A SS training scenario. In general, mechanical solutions to haptic feedback are either 3D printed or molded, offering limited scalability, compared to electrical approaches. However, physical shapes are more reliable: for instance, discrimination performance of tactile contour images on TeslaTouch was close to chance level [158].

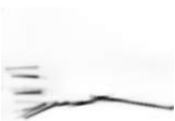

Similarly to haptic feedback, Maidenbaum, Abboud and Amedi [142] mentioned the ignorance of accounting for the sensory-motor loop in training protocols. Robust, fast perceptual learning likely involves the dependence of SS stimulus on motor movements: e.g. alteration of soundscapes during head, or hand movements when the visual field encompasses the hands of the user. The visual reference frame should also be constrained by the application of a head-mounted camera, instead of allowing the blind to reposition it arbitrarily. For instance, smart glasses or a cheaper Google Cardboard can overcome this constraint.

Differences between explicit and implicit conversion methods emerge in SS training as well. The work of Proulx and colleagues [13] argues that if the substituting stimuli shares spatiotemporal attributes with the corresponding tactile or motoric feedback, then that implicit correlation is sufficient to induce perceptual learning and generalization. Furthermore, implicit training suffices for SS of simple, abstract shapes [32, 162]. Even though the conversion method used in these experiments is explicit, the V2A logic was not explicitly described to the subjects. However, discrimination of complex visual objects and scenes are more difficult to generalize, when no explicit instructions are provided: for example, subjects incorrectly associated semantic attributes to auditory features in the study of Kim and Zatorre [32]. Another complex task, reading characters using TV, also required explicit teaching [33].

A major setback in the expansion of V2A SS devices is the absence of organized training procedures [142, 165]. In an ideal case, the visually impaired would receive supervised training, interacting with experts, combined with portable support materials [155]. Complexity of the visual stimulus should be gradually increased, starting with simple, geometric shapes [9], continuing with dynamic stimuli and finally rich, real world tasks to perform [142]. The higher the variety of exercises, the more robust cross-modal associations become by transfer learning. Additional gamified elements boosts motivation and grants the feeling of accomplishment, which is necessitated by the months of demanding training [155]. Finally, appropriate electrical or magnetic stimulation (TMS) of visual regions may hasten perceptual learning of V2A SS stimuli, by guiding cross-modal synaptic plasticity [157, 166].

## 2.6 Deep learning

### 2.6.1 Deep neuroscience

Initially, neuroscience aided the design of deep learning models; now deep learning has turned full circle and provide modeling tools to computational neuroscience. On an abstract level, the brain is a recurrent neural network of great depth, abundant of nonlinearities. The shallow, hand-crafted models of neuroscience are incapable to scale and simulate such networks, solely yielding low-level tuning functions [167]. As DNN architectures and training procedures become more structurally complex and specialized, they begin to share a wider conceptual ground with neuroscience. Such conceptual commonalities include the process of attention, recursion and memory consolidation [168].

To complement the detailed, yet shallow bottom-up models of computational neuroscience, top-down solutions have been developed in the form of DNNs, which succeeded in mimicking various biological responses. Kell and others [169] generated, optimized and compared 200 hierarchical deep learning architectures performing a speech and music recognition task. In their study, CNNs were trained to extract categorical information of music and speech from cochleagrams. The trained CNNs achieved human-level accuracy and further exhibited human-like errors in their classification. Moreover, layers of the network predicted fMRI BOLD patterns in AC, more precisely than classical spectrotemporal filter models.

The visual cortex holds the highest track record of being emulated by DNNs: Yamins and colleagues [170] designed a CNN that resembles the activation of the human inferior temporal cortex, the highest ventral visual cortical area, when encoding complex naturalistic images. Intermediate layers also showed correlation to V4 responses, which network provides input to the inferior temporal areas. Similarly, Kuzovkin and others [171] showed parallel between the frequency of neural activations and CNN layer representations of images.

In a study of mammalian navigation, a long short-term memory (LSTM) network was able to realize grid-cell-like behavior when trained on a path-integration task. The learning model received velocity vectors as input, similar to what the mammalian brain operates on while maneuvering [172]. Bhalla [173] further pointed out that LSTM networks mirror the dendritic sequence discrimination process in the hippocampus, with the inner gates of the LSTM cell roughly resembling regions of the hippocampus in their functional, recursive connections.

Even though, DNNs abstract away a huge body of details from biological neural networks, if we are able to align the learning objectives and neural structures operating in the brain to closely equivalent deep learning models, the two fields of study, neuroscience and deep learning, could aid one another in testing neuroscience hypothesis in silico, and in establishing biologically inspired computational optimization methods.

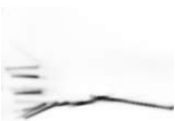

### 2.6.2 Autoencoders

Discovering a fitting conversion method from images to sound can be defined as an optimization problem. Optimal conversion in this context refers to a compression mechanism that acquires the shortest sequence of sound samples with minimal information loss given the image to encode.

Autoencoder models are well-known for their exceptional ability to generate compressed encodings by massively employing nonlinear transformations. The simplest autoencoders are end-to-end trained by gradient descent, the input and desired output of the network being the same. Such learning models consist of an encoder and a decoder network, with a latent bottleneck in between (Figure 11). The degree of freedom of the bottleneck determines the ensuing compression: the smallest layer in the model, often placed in the middle, is forced to squeeze the input information, so the decoder can reconstruct the input out of the compressed representation. In case of V2A substitution, the bottleneck is the audio representation, while the model is fed with the same image as the output we expect from it. Recurrent autoencoders merely involve recurrent neural layers as part of the encoder and/or the decoder.

Variational autoencoders (VAE) differ from regular autoencoders by the way in which they distribute the latent, compressed representation [174]. The latent space is forced to spread in a Normal distribution, and as a result, similar input representations are mapped closely together in that space, which constructs the space to be more intuitive to walk through: a slight change in a latent variable corresponds to a small and consistent shift of certain characteristics in the decoded representation. Hence, a VAE is a better fit for the V2A SS problem, as the latent space, i.e. the distribution of soundscapes, becomes more intuitive, easier to understand and to learn by the blind, compared to a regular autoencoder. Moreover, by expanding the latent space to a Normal distribution, we require the model to generate diverse audio stimuli, so distinct visual features are expected to be encoded as perceptually distant sound sequences.

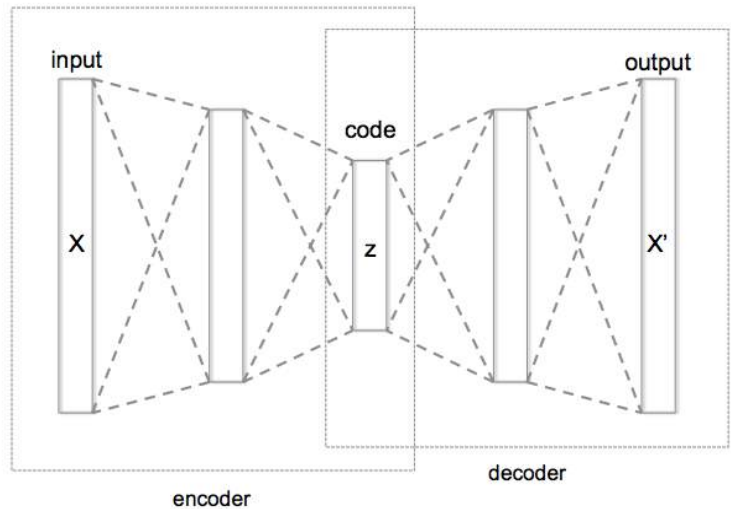

Figure 11: Depiction of an autoencoder. Autoencoder models are trained to reconstruct the input image, $X$, in their output $X'$, after compressing it to a latent representation, $z$.

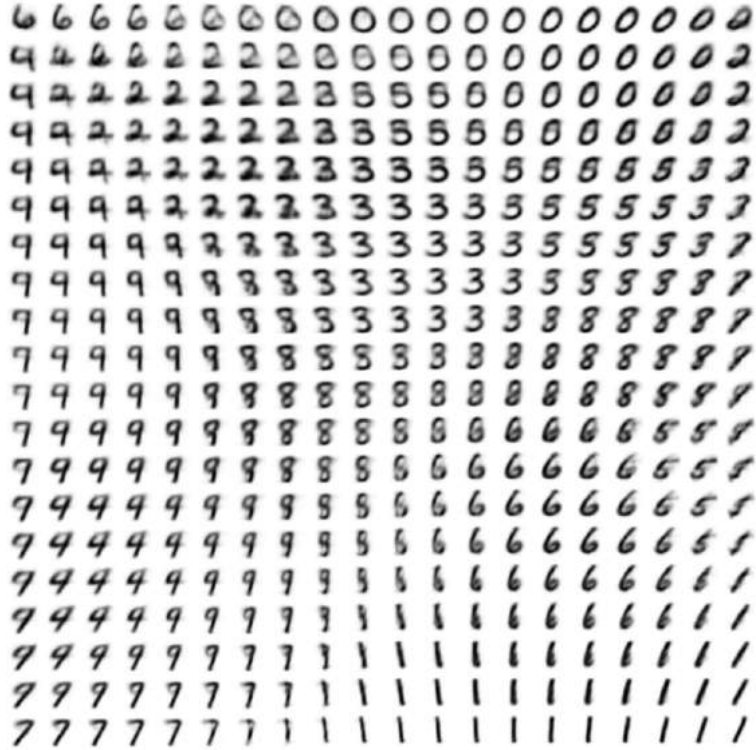

Figure 12: VAE learned manifold of the MNIST dataset. A VAE with a two-dimensional latent space was trained on handwritten numbers; this figure depicts the output images of the decoder as the latent space is scanned through. The hidden representation managed to incorporate an intuitive interpolation between visual features of the written numbers. Reprinted from the work of Kingma and Welling [174].

Disentangled VAEs [175] are slightly modified versions of the variational models: the independence of latent variables is imposed stronger, which leads to an even more intuitive compressed representation; albeit, occasionally to the detriment of the reconstruction quality.

The Deep Recurrent Attentive Writer (DRAW) model is a fine instance of a recurrent VAE [35]; it is also used as the foundation for the AEV2A model presented in this study. The authors of DRAW pointed out that drawing is a sequential process, corrections are made to the image by repeatedly reassessing the canvas before the next modification. Their model reflects this insight (Figure 13): the encoder reads parts of the input image and the unfinished drawing on the canvas, then compresses the information into a latent vector. The decoder additively writes on the same canvas given the latent variables. The encoder further receives the hidden state of the decoder, i.e. an impression of previously emitted modifications. Both the encoder and decoder are implemented as LSTM networks. DRAW applies a particular reading and writing attention mechanism: in each iteration, a grid of two-dimensional Gaussian filters are positioned on the image and the canvas. The isotropic variance of the

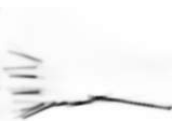

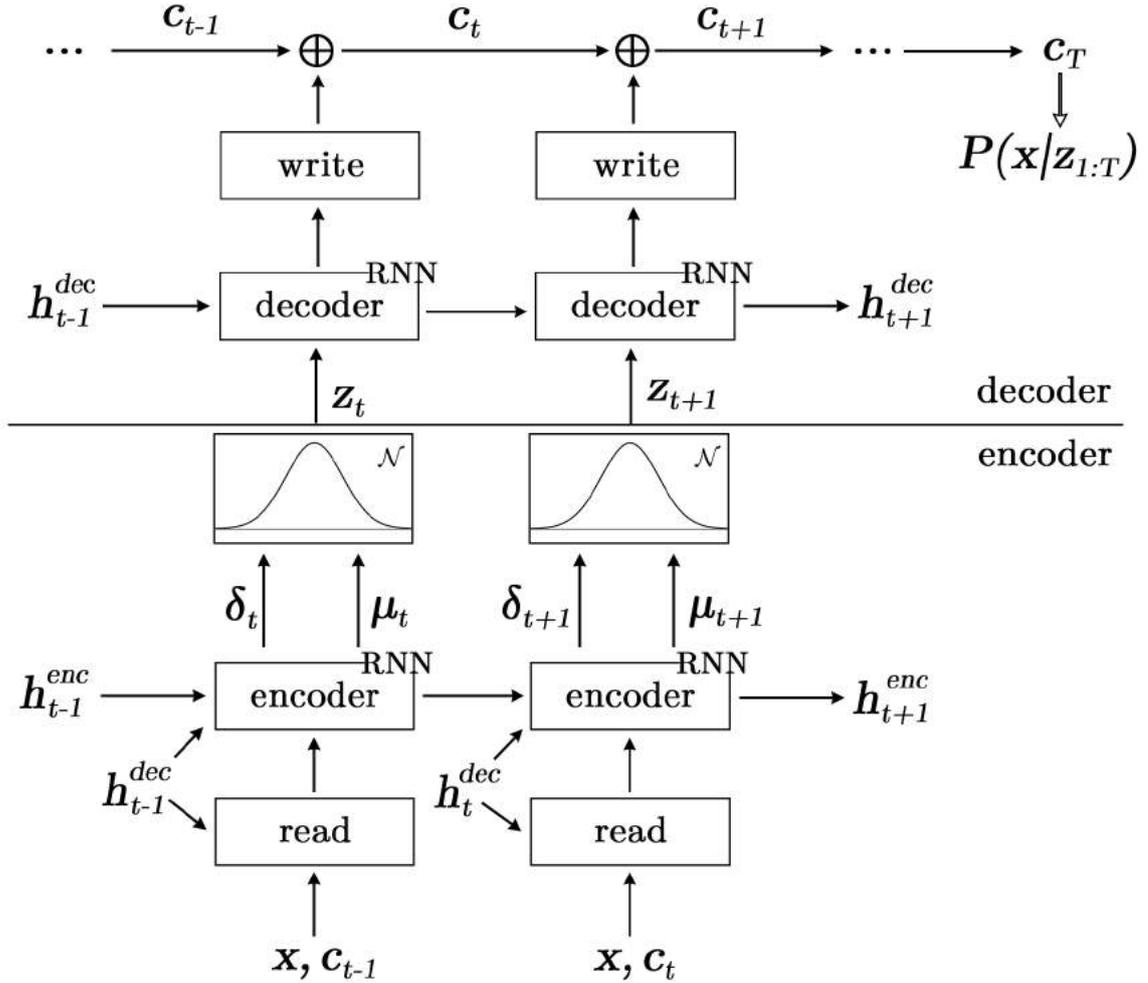

Figure 13: The architecture of the DRAW model [35], unfolded for two recurrent iterations. After $T$ iterations of drawing on the canvas, $c$, the network is optimized to reconstruct the input image, $x$. The reader attention module receives $x$, the previous state of the canvas, $c_{t-1}$, and the previous hidden state of the decoder, $h_{t-1}^{dec}$. Hence, reading attention is allocated according to what has been written until $t$. The latent states, $z_t$, are random drawn from Normal distributions parametrized by the recurrent encoder unit. Both reading and writing is performed by a grid of Gaussian patches.

Gaussian, the stride distance between neighboring filters and the coordinates of the grid center are dynamically determined by the network, having the freedom to decide the location and detail of the reading and writing attention. An additional intensity variable defines the strength of the filter response. The reconstruction loss of DRAW is specified as the L2 distance between the final state of the canvas and the input image.

Autoencoder models can be further extended by a discrete bottleneck. For instance, Oord, Vinyals and Kavukcuoglu [176] substituted the variational aspect with a vector quantization method (VQ-VAE) that wields the encoder with a discrete output and lets it learn the prior instead of relying on a predefined one. A discrete latent representation yields two major benefits: 1) it counteracts the posterior collapse VAEs are typically suffering from, which causes a model with a strong, deep decoder to ignore the latent representation, driving the network to an undesirable local optima, and 2) it controls the information content directly by the amount of discrete vector code space available. The latter advantage is critical in the design of SS conversion logic, as substantial information content may be compressed even in a single continuous variable [177], and thus the encoded sound stimuli of distinct visual features would barely be distinguishable. Dieleman and others [178] further improved on the discrete bottleneck, proposing the argmax autoencoder, which slightly underperforms the VQ-VAE, yet converges more reliably.

From the marriage of VAEs and Generative Adverserial Networks (GAN) [179], the adverserial autoencoders (AAE) was born [180]. AAEs spread the latent space more evenly compared to VAEs, causing the model to generalize better and be more faithful to the imposed prior distribution. Ideally, if we could define the distribution of perceivable and distinguishable soundscapes, it could be enforced as a prior, in which case, AAE solutions would have an advantage.

When the design of implicit V2A SS conversion methods is considered as a compression problem, autoencoders and their variants come naturally to the rescue. The combination of highly nonlinear compression and diverse, intuitive latent encoding, enables us to reliably convert a predefined set of images to sound and back. When matched with explicit SS conversion logic, a drawback to this approach is the inability of the network to encode visual information that is outlandish in the face of the trained image set. Nevertheless, the training set having an inherent structure, the autoencoder can compress on it, and realize sound representations that are shorter, yet perceptually still discernible.

### 2.6.3 Audio coding

Historically, audio signals have been much more difficult to synthesize and to decode compared to images. This difficulty stems from the long-range temporal structure of audio: from spectral information stretching over 100–1000s of samples, to rhythm spanning over the timescale of seconds or 10s of thousands of samples. Compression and generation of high quality sound signals in a DNN setting have only been solved in recent years [115].

Classical approaches of audio encoding extract the spectral power of the sound in short temporal windows, ignoring the phase information, rendering the obtained representation non-invertible. Moreover, as we described in Section 2.3.3, phasic cues are critical to pitch perception, thus generated audio from spectrograms lack substantial bandwidth of information. The derived spectrograms are usually further processed by two-dimensional CNNs, for instance, the SpecGAN model of Donahue, McAuley and Puckette [181] does so. On the other hand, WaveGAN, also presented in the same study, employs one-dimensional convolutional filters to raw audio samples, and performs better on subjective sound quality measurements than SpecGAN.

Hitherto, deep autoregressive models have been the most successful in sound encoding. Autoregression implies the involvement of past filtered values in the computation of the predicted value, meaning that the output of the model at time $t$ is required to estimate the output at time $t + 1$. For example, IIR filters can be viewed as autoregressive, as they rely on previous outputs, while a FIR filter is merely a weighted moving average of input values.

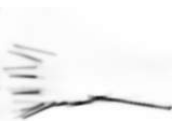

WaveNet [115], a deep autoregressive neural network, applies layers of dilated causal convolutional filters. The dilated nature of its filters realizes an exponential growth of the temporal receptive field with more layers appended, while saving a great quantity of parameters compared to regular, stacked convolutions. The causal aspect, not present in WaveGAN for example, enforces WaveNet to derive the output by only incorporating past input states, emphasizing the temporal characteristic of audio. Being an autoregressive model, WaveNet performs sequential predictions, which is significantly slower than the parallel sound synthesizer competitors. The later developed, sped up version of the model, spends more than 50 seconds to generate 1 second worth of 16000 samples [182]. Another spin on the model [183], however, trains a parallel network taught by an already trained WaveNet, achieving 20 times faster than real-time speech generation. Yet another fine example of an autoregressive autoencoder is the argmax autoencoder [178], which yields slightly higher scores on human evaluations than WaveNet.

If we aim to synthesize and/or decode sound in an implicit V2A SS conversion function, we need to consider the above mentioned DNN models, particularly the autoregressive ones. However, none of these DNNs parallel the representational power of the human auditory system and still they demand substantial computing power and time to train. They have mostly been tested with a low sample frequency of 16000 Hz, and very constrained sound space, like speech signals of a given language. Due to such downsides, we need to resort to hand-crafted synthesizers and simplified, psychoacoustic-based audio decoders; at least for the time being.

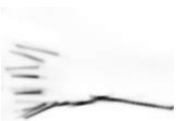

# 3 Methods

## 3.1 Autoencoded Vision to Audition (AEV2A)

Ultimately, the aim of this study is to realize an encoder that converts images, a high spatial, low temporal dimension representation, to audio, a low spatial, but high temporal dimension space. Previously, V2A SS devices have employed an explicit conversion logic, which assigns low-level visual features, like pixels, to predefined sound sequences. Such devices are based on multisensory processing research to explicitly define visual–auditory feature correspondence: 1) higher pixel position is translated to higher pitch, 2) higher luminance is converted to higher amplitude, and 3) pixel position is further conveyed by sound source location. Previous SS solutions demands months of training to achieve beneficial use of the gained sense, and years to acquire synaesthetic visual perception [10].

The slow learning rate of such SS devices prevent the large population of visually impaired to take advantage of them. We hypothesize two principal reasons for the relatively slow learning rate: low frame rate of the devices, and the ignorance of human auditory compression. A third major reason that has been proposed before [142, 165], is the lack of well designed SS training programs outside the lab. The latter issue is only touched on in this study.

As autoencoders are designed to realize a nonlinear compression function, it fits to be trained as an implicit V2A conversion logic. The proposed Autoencoded Vision to Audition (AEV2A) model involves an abundance of hyperparameters, which can be fine-tuned to remedy the above mentioned reasons for the slow learning rate. The hyperparameters include, the image set the model is trained on, the level of visual space abstraction, the properties of the synthesized audio and the degree of visual–auditory feature correspondence. The TensorFlow implementation of the model can be accessed from `github.com/csiki/v2a`.

The high dimension of an image can be further decreased by extracting the contour of the image using a simplified computational model of V1 simple cells; i.e. edge detection. Lowering the spatial dimension in such a way is loosely built on the research of metamodal sensory computation, which has found that the unused sensory brain areas are acquired by other active sensory areas for ease of computation. In our case, the unused area is V1, the active sensory region is A1. The visual pathway preceding V1, which is supposed to perform contour detection, suffers serious apathy in the blind. We extract the contour as a preprocessing step, to overcome due to the blind's lack of functioning brain areas that could execute such a computation. We cross-modally cater information content for V1, similarly to which it naturally receives in the sighted. We hypothesize that this setup demands less adaptation or neural plasticity from the user, and thus could hasten SS training. Moreover, the contour of the image is the closest visual representation that shares the majority of its features with haptic sensing, accelerating SS training according to the body of research on perceptual learning.

The audio synthesizer is designed to elicit responses in A1, instead of relying on low-level audible features, like raw frequency content, which are typically encoded earlier in the auditory pathway. This design incorporates cross-modal studies of the blind, showing that cortico–cortical connections are likely wired between A1 and V1, the recruited V1 only being able to process representations present on the level of AC. The hyperparameter set of the audio synthesizer can be further adjusted to comply with the capabilities of human hearing, documented in psychoacoustic studies.

### 3.1.1 Model structure

The Autoencoded V2A (AEV2A) model builds on the recurrent VAE architecture of DRAW [35]. The model consists of an encoder and a decoder half, both parts include multiple layers of LSTM cells. Figure 15 shows the AEV2A unrolled for two iterations, $t$ and $t + 1$.

Similarly to DRAW, the input image is first read by an attention unit, applying a grid of Gaussian patches with parameters like isotropic variance, stride and grid location determined by the network in each iteration. The reader attention parameters are learned over training and derived from the input image, the current state of the canvas, and the previous hidden state of the decoder LSTM. If the grid contains $N \times N$ patches, $N^2$ values are passed to the encoder LSTM, which outputs the variance and mean of the Gaussian distributions used to randomly draw the hidden states from. The drawn latent values parametrize the audio synthesizer, generating a soundscape representative of the visual information obtained in that iteration.

The decoder half either receives the generated audio stream, or a rescaled variant of the input parameters the synthesizer gets, the rescaling of which reflects human hearing limitations. The former is termed as the hearing decoder, the latter as the deaf decoder. The distinction between the two is elaborated in Section 2.3.6. A set of computational hearing models compress and/or noise the received sound representation, which is then consumed by the decoder LSTM. The output tensor of the recurrent network is fed to the writer attention unit. The writer attention either applies the same grid of Gaussian patches as the reader attention, or a collection of edges, each of them delineated by a bundle of Gaussian patches, with the same parameters, but distributed in a line instead of a square grid. The latter is termed as V1 attention, as it draws edges on the canvas analogous to the visual features simple cells in the occipital cortex are tuned to.

The canvas is updated in each iteration, the final state of which should depict the original input image, i.e. the distribution of which should follow the distribution of the input images given the sequence of latent variables. We define the number of total iterations as constant.

The multilayered decoder LSTM implements residual, skip connections between non-consecutive cells, which results in smoother clustering of latent variables and reduces the likeliness of posterior collapse [184].

The following set of hyperparameters define the network structure; typical values that were experimentally tested are shown as well:

– Sequence length: the number of iterations performed; up to 40

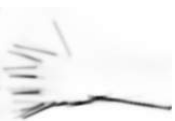

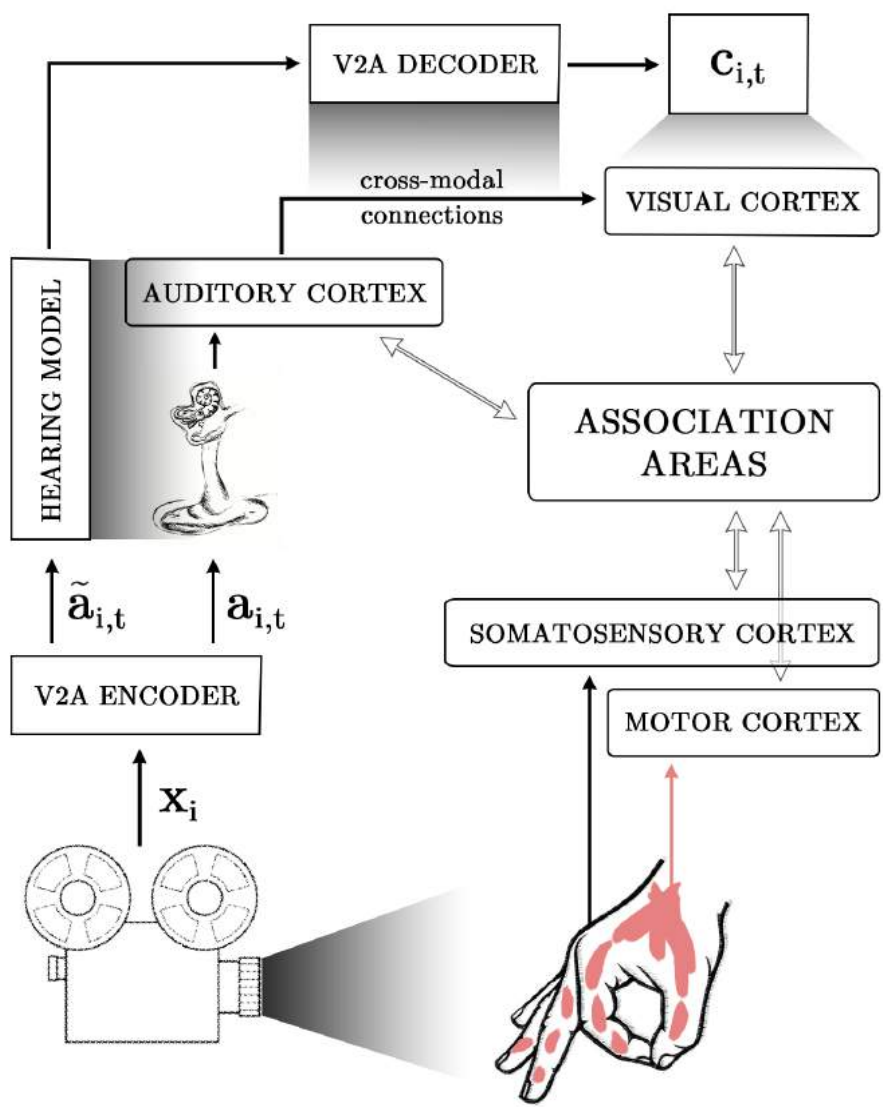

Figure 14: In V2A SS, visual information is conveyed through sound to aid the blind. AEV2A applies a deep recurrent autoencoder model to synthesize sound, $a_{i,t}$, from images, $x_i$, decoding the produced sound back as a sequence of drawing operations on a canvas $c_{i,t}$. The human hearing model included in the autoencoder aims to constrain the synthesized audio to be perceivable by human listeners, mimicking the auditory pathway. The canvas loosely emulates the visual cortex, as the drawing operations are performed in units of features that the simple cells are tuned to. When the visually impaired users wear such a V2A SS device, and e.g. listen to the image of their hands, they integrate the muscle and haptic information with the heard audio, and learn the correspondence between these sensory features by unconsciously adjusting to the feedback signals from higher multisensory and associative areas. Roughly speaking, AEV2A learns to encode visual information in a sequence of soundscapes, constrained by human hearing, so they can be decoded as successive drawings on V1.

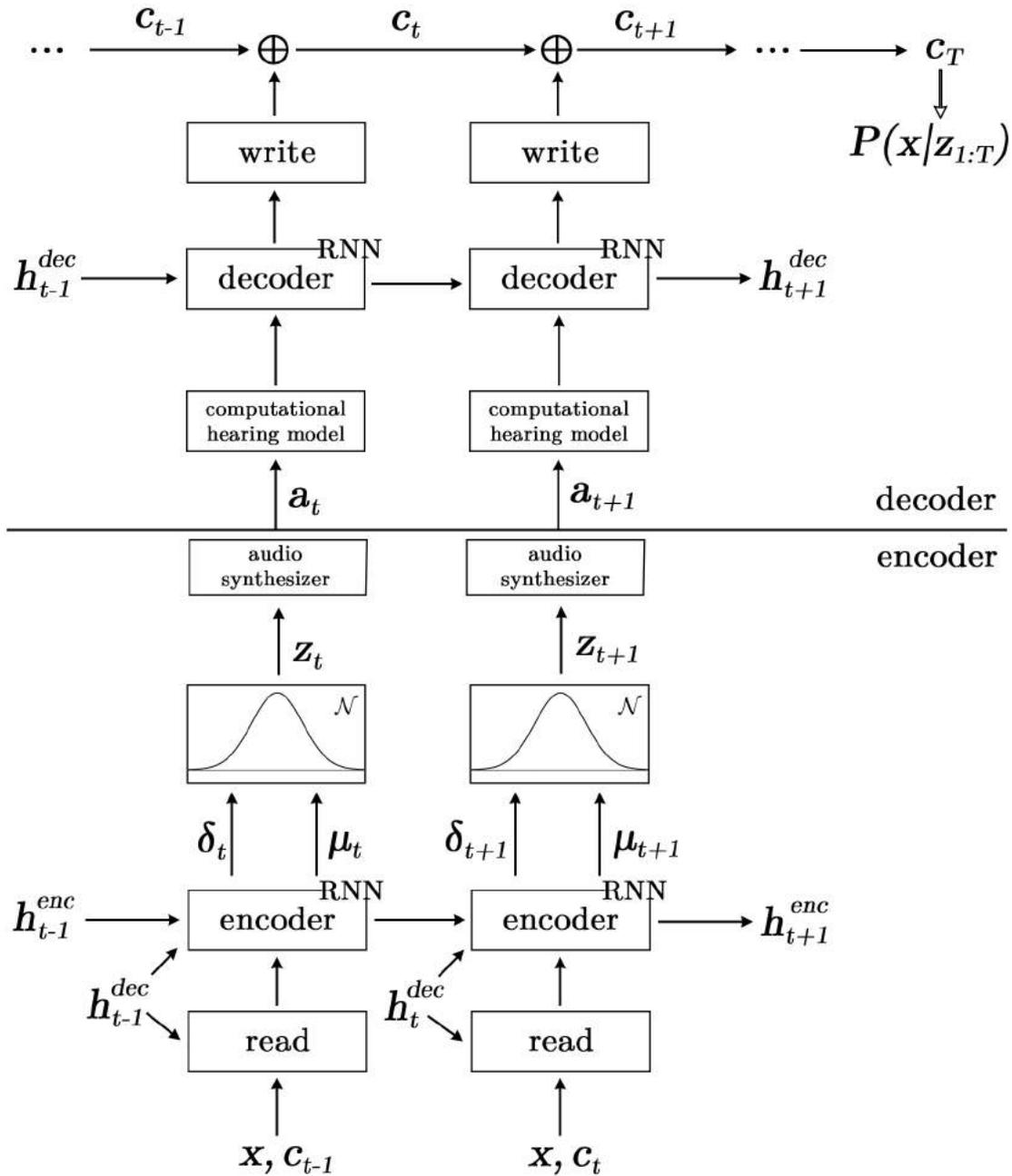

Figure 15: Unrolled architecture of AEV2A. Compared to the DRAW model (Figure 13), AEV2A includes a hand-crafted audio synthesizer and a hearing model. Both of these extensions aspire to shape the latent sound representation to comply with human auditory coding. The synthesizer is constructed so different input parameters, $z_t$, yield distinct neural activations in A1, while hearing unit should compress or noise away any detail that is not humanly perceivable.

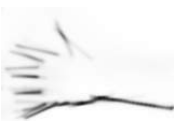

– Number of LSTM layers in the encoder: 3 or more

– Number of LSTM layers in the decoder: 3 or 4; should not be high, as that may lead to posterior collapse; furthermore, we should not assume the existence of a complex sound to visual information decoding network in the human brain, or analogously, we should presume that cross-modal connections between the auditory and visual areas, which the decoder LSTM aims to roughly emulate, are scarce

– Dimensionality of LSTM cells: 512 or 1024; higher values yielded no gain

– Dimensionality of the latent space: either the number of parameters the audio synthesizer requires or more; if more, a dense layer is inserted between the latent state and the synthesizer

### 3.1.2 Visual features

Input images undergo three preprocessing steps before they are fed to AEV2A: 1) they are downsampled to the size of $160 \times 120$, 2) turned to a grayscale representation, and 3) their edges are extracted.

The contour detection algorithm of choice is the push-pull CORF method [71], due to its faithful modeling of V1 simple cells and its high signal-to-noise ratio. The maximum of the CORF generated simple cell responses are taken over the orientation dimension to achieve a smooth two-dimensional image of edges. Although, the described process achieves visual space abstraction to the features of bars, those bars are still embedded in two-dimensional image space, instead of being explicitly described by their positions and orientations. From the perspective of the learning algorithm, encoding the contour poses a simpler problem than translating the whole image to soundscapes, as the visual space shrinks significantly.

The parameters given to the push-pull CORF algorithm are the following: sigma is set to 2.2, beta to 4, the inhibitor factor is defined as 1.8. The Matlab implementation of the push-pull CORF model used in this study can be accessed at Mathworks [72]. We made a slightly optimized version, which can be retrieved from the AEV2A repository: github.com/csiki/v2a.

### 3.1.3 Audio synthesis

The AEV2A model relies on the cross-modal recruitment of the occipital cortex for the processing of auditory stimuli, attaining an additional computational power, desperately needed in the V2A SS domain. To maximize the likelihood of cross-modal recruitment, we designed a sound generator algorithm that explicitly synthesize auditory streams; three arguments support our design: 1) the cross-modal connections appear to stem from A1 [39], 2) A1 encodes higher-order stimulus qualities, such as AM and FM, more accurately than low-level features such as pitch or amplitude [20, 21, 23–25], and 3) V1 responses correlate with higher-order processing of auditory streams instead of low-level sound discrimination already performed in the auditory system [33, 102]. By compiling these arguments, we hypothesize that higher-level auditory features of AM, FM and spatial modulation (SM) of sound, composed into soundstreams, are suitable SS stimuli, if cross-modal recruitment of the striate cortex is preferred. The case of SM is further strengthened by the study of Poirier and colleagues [119], demonstrating that in the blind, auditory motion processing is likely to be partially transferred to occipital brain networks, which are responsible for visual motion processing in the sighted.

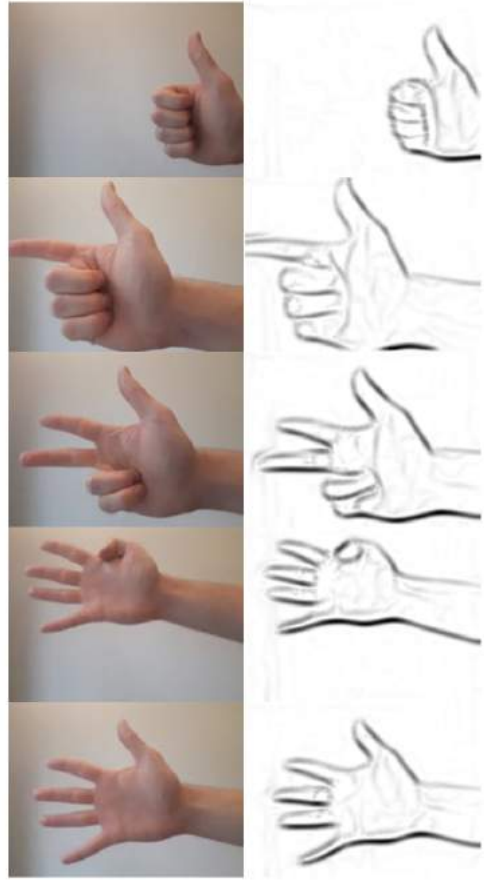

Figure 16: Images from our hand posture dataset and their corresponding contour representation. Edge detection is performed by the push-pull CORF algorithm.

The dimension of sound elevation is not included in the SM of sound, only the modulation of azimuth. As sound localization accuracy varies by elevation in a nonlinear fashion [119], setting the elevation to zero simplifies our grip on the distribution of the localization error, which is paramount to our binaural hearing model described in the next section. Furthermore, elevation discrimination accuracy is fairly inferior to azimuth encoding in humans [96].

In the AEV2A model, there are two major ways to enforce the correspondence between dissimilar visual features and perceptually distinct synthesized sound stimuli, i.e. to generate soundscapes we can distinguish between. Either the encoder is constrained to synthesize such audio, and/or the decoder filters out, compresses away auditory features that are not reliably encoded in the human AC. The latter, termed the hearing decoder as opposed to the deaf decoder, would likely require a computational model of human hearing to be embedded in the learning model, the gradient of which is necessary to be specified. We elaborate on the strengths and weaknesses of this approach in the next section.

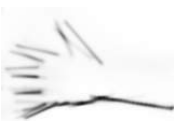

The hyperparameters of the proposed synthesizer can be adjusted to comply with the limitations of human hearing, and produce sounds that are perceptually distinguishable. Namely, we can pressure the synthesizer to output perceptually different soundscapes in response to different input variables, relying on psychoacoustic research. Effectively, we consider the input variables as addresses to the spectrum of auditory perceptual states. The synthesizer first translates the input to units of pressure amplitude, frequency and interaural cues, by applying the inverse of the tuning function specified for the sense of loudness, pitch and observed azimuth, respectively. For instance, loudness is perceived logarithmically, similarly to frequency, therefore, their corresponding variables are scaled exponentially.

To convert loudness to amplitude, we employ the inverse function of loudness perception. Inverse of $L = A^{\alpha_A}$ is $A = L^{\frac{1}{\alpha_A}}$, where $\alpha_A$ is set to 0.3 [76]. To derive frequency, we handle the input variable, $p$, as an address on BM; we translate it to the corresponding CF: $f = \lambda_f(10^{\alpha_f * p} - k_f)$, where, $k_f = 0.85$, $\lambda_f = 165.4$, $\alpha_f = 2.1$ [84]. $\lambda_f$ may be decreased to avoid annoyingly high frequencies in our soundstreams. For example, setting it to 92, yields a maximum frequency of 11603 Hz, a value of 60 caps the frequency at 7600 Hz.

Given the azimuth angle, $\theta$, ITD is delivered according to the Woordworth model [185]: $ITD = \frac{r}{c}(\theta + \sin\theta)$, where $r$, the head radius, is set to 8.75 cm, $c$ is the speed of sound, and $\theta \in [-\frac{\pi}{2}, \frac{\pi}{2}]$. We constructed our own frequency-dependent ILD function, by fitting ILD lines of different frequencies to Figure 1 of Ref. [94]: $ILD = 20 * (\frac{f}{10000})^{0.42} * \sin\theta$ in dB. This model does not reflect the ILD bump present at 1000 Hz, neither the asymmetry of ILD as the frequency increases. It was designed to reach 20 dB of ILD at 10000 Hz and 90 degrees of azimuth, and to closely fit ILD values at 250, 500, 750 and 1000 Hz, for sound sources at 0.5 m.

We synthesize a stereo soundstream as summarized by the following equations:

$$s_t^{\mathrm{L}} = A_t * \sin\left(f_t * 2 * \pi * t + ITD_t^{\mathrm{L}}\right) * ILD_t^{\mathrm{L}}, \tag{4}$$

$$s_t^{\mathrm{R}} = A_t * \sin\left(f_t * 2 * \pi * t + ITD_t^{\mathrm{R}}\right) * ILD_t^{\mathrm{R}}, \tag{5}$$

where $t \in [1..S]$ is the audio sample index, $A_t$ represents AM, $f_t$ determines FM, while $ITD_t$ and $ILD_t$ realize the binaural cues of SM. A soundstream consists of sections, the number of which, $M$, and the time between sections, $T$, are constant within a AEV2A model. The soundstream sample length, $S$, is hence defined as: $S = M * T * f_{\mathrm{s}}$, where $f_{\mathrm{s}}$ is the sample frequency.

Input tensors fed to the synthesizer are separated into ground values and modulation vectors. Ground values specify the offsets for amplitude, frequency and azimuth, while modulation vectors detail how these audio qualities change over time within a stream. Ground values may be viewed as absolute, while modulation vectors as relative, successive shifts. Separation of these parameters, in contrast to providing merely an array of absolute values, allows us to restrict the AM, FM and SM modulation rates directly. Tolerating too high rates, i.e. steep shifts in the qualities of sound, could lead to fissure [101], so a single soundstream is inadvertently divided into multiple auditory streams, perceptually. Humans are able to discriminate shapes of AM rates of 100 Hz [24] and are sensitive to FM rates of 64 octave/sec [25], which roughly translates to $T$ being around 10 ms, with an upper limit of 0.64 octave FM modulations under $T$.

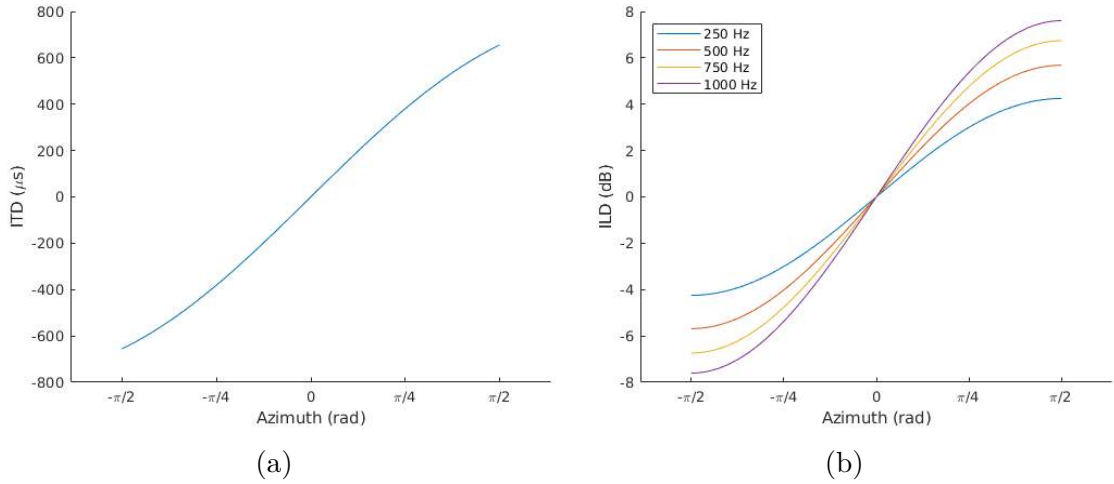

(a)    (b)

Figure 17: ITD and ILD by sound location azimuth. ITD calculated according to the Woodworth formula, ILD is based on our frequency-dependent, simple, fitted model.

The input modulation tensors, $L_m$, $p_m$, $\theta_m$, of length $M$, delineate the degree of amplitude, frequency and azimuth change between sections; $L_0$, $p_0$ and $\theta_0$ represent the offset. To derive the final modulation vectors used in Equations (4), and (5), we first compute the cumulative sum of the modulations, adding the offset value to each element:

$$L'_m = \sum_{i=1}^{m} L_i + L_0, \tag{6}$$

$$p'_m = \sum_{i=1}^{m} p_i + p_0, \tag{7}$$

$$\theta'_m = \sum_{i=1}^{m} \theta_i + \theta_0, \tag{8}$$

where $m \in [1..M]$, $L'_m \in [0,1]$, $p'_m \in [0,1]$ and $\theta'_m \in [-1,1]$.

Second, we apply the human inverse tuning functions described above, so distinct values would likely correspond to distinct perceptual states:

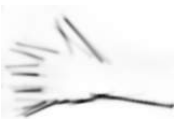

$$\tilde{A}_m = (L'_m)^{\frac{1}{\alpha_A}}, \tag{9}$$

$$\tilde{f}_m = \lambda_f(10^{\alpha_f * p'_m} - k_f), \tag{10}$$

$$\tilde{\theta}_m = \theta'_m * \frac{\pi}{2}. \tag{11}$$

Finally, vectors, now incorporating the ground and modulation values, are linearly upsampled to length $S$, leading to the tensors $A_t$, $f_t$ and $\theta_t$. This upsampling results in a linear change of sound qualities between modulation states; i.e. sections implement a smooth transition between the audio states specified in the input.

$\theta_t$, representing the sound source azimuth, is further translated to ITD and ILD cues by employing the Woodworth model [185] and our own, custom fitted, frequency-dependent ILD function:

$$ITD_t = \frac{r}{c}(\theta_t + \sin\theta_t), \tag{12}$$

$$ILD_t = 10^{(\frac{f_t}{10000})^{0.42}} * \sin\theta_t, \tag{13}$$

where $ILD_t$ is defined in sound pressure ratio of the left and right channels.

We equally influence the left and right stereo channels to implement the time and level differences, thus, our channel specific variables are the following:

$$ITD_t^L = \frac{ITD_t}{2}, \tag{14}$$

$$ITD_t^R = -\frac{ITD_t}{2}, \tag{15}$$

$$ILD_t^L = \sqrt{ILD_t}, \tag{16}$$

$$ILD_t^R = \frac{1}{\sqrt{ILD_t}}. \tag{17}$$

Multiple soundstreams are overlayed to compose a soundscape. Given the length of the soundscape, the enclosed soundstreams are spread equally within it, so consecutive streams have the most overlap. Although, we have no mechanism in place to assess the accuracy by which overlayed soundstreams are encoded in AC, soundscapes of different stream composition should elicit various timbre characteristics. Ideally, the temporal offset of streams within a soundscape should be part of the trainable parameters in the network, but we have not found a solution to derive the gradient with respect to such a temporal delay.

The introduced audio synthesizer generates audio incredibly fast, regardless of sampling rate. Out of a small number of input parameters, tens of thousands of samples are spawned in parallel.

The synthesized audio is also relatively pleasant to the ear. All the soundstreams are ramped up and down for some milliseconds to avoid spectral splatter. The output is akin to a group of diverse birds chirping simultaneously, or to fast paced sci-fi computer beeps.

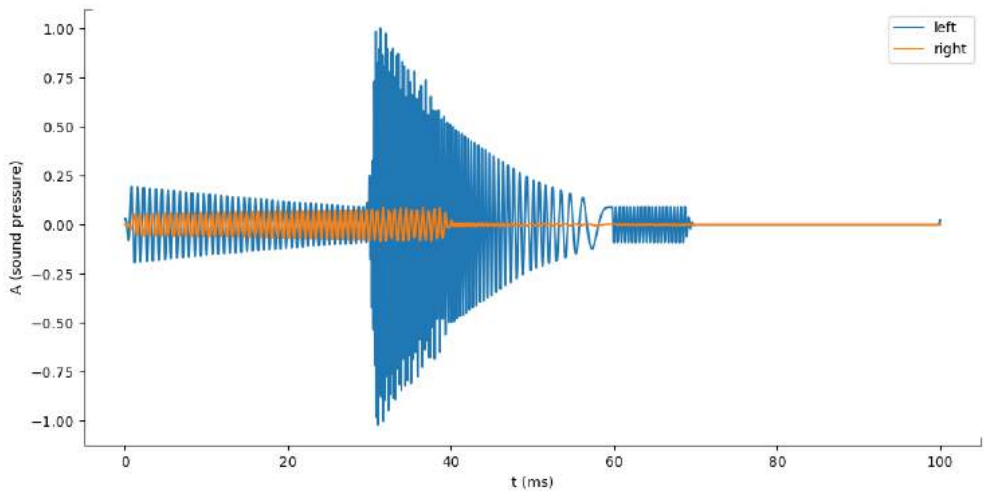

Figure 18: A random generated soundscape of three soundstreams with 25% overlap and 4 sections each, every section being 10 ms long. The first stream manifests a slowly damping amplitude, while the azimuth transitions from left to right. The frequency of the second soundstream decreases over time and played predominantly in the left channel. The third stream has a close to zero amplitude.

The synthesizer relies on the following set of hyperparameters; typical values that were experimentally tested are shown as well:

– Sampling frequency: number of audio samples per second; 44100 used for deaf decoders, and 16000 or 22050 Hz for hearing decoders

– Number of soundstreams in a soundscape: 2, 3 or 4, if more is included, the soundscape becomes convoluted, depending also on the amount of overlap

– Number of sections: number of modulation states in a soundstream; at least 3

– Section length: time between modulation states; at least 5 ms, less than 12 ms, 10 ms is tested the most

– Ratio of soundscape length to the soundstream duration: implies the overlap between streams; e.g. if 2, and the number of streams is set to 2 too, then there is no overlap; 1.5, with again 2 soundstreams enclosed, results in a 50% overlap; depends on the number of soundstreams included, but mostly experimented with values between 1.2 and 2

– Attack and decay length of soundstreams: the duration of ramp up and down for each stream; as low as 1 ms of each eliminates spectral splatter

### 3.1.4 Hearing models

In AEV2A models, we differentiate between two forms of decoding: hearing and deaf. The deaf decoder receives the variables $\tilde{A}_m$, $\tilde{f}_m$ raw (Equations 9, 10), and reconstructs the drawing patterns without needing to extract features from the synthesized sound. On the other hand, the hearing decoder receives the audio samples, and essentially needs to distill the synthesizer variables that the deaf decoder takes for granted. The hearing decoder, hence the name, may include computational models of human hearing to control for nonlinear compression processes, such as distortion products, or spectral masking. By including this compressive bottleneck, we filter out audio information that the human auditory pathway would not encode, i.e. that we would fail to perceive. Computational hearing models expect soundwaves as input, which prevents deaf decoders to implement these hearing models.

Hearing decoders are orders of magnitude slower to train, because their computational demands scale with the number of audio samples, $S$, instead of the number of modulations, $M$. Hence, sampling frequency often needs to be decreased from 44100 to 22050 or even to 16000 Hz, which restricts the frequency range of the audio. Furthermore, extracting auditory features in DNNs still poses serious difficulties: sound decoder models are either heavy to train [115], or limited in their efficacy to support diverse audio signals [181]; either way, they are far inferior to the abilities of the human auditory system.

Although we synthesize stereophonic sound, the hearing decoder only receives the monaural version absent from ITD and ILD cues. Doing so, the necessary computational resources are halved. Moreover, binaural hearing models are quite complicated and would further decelerate the training procedure.

Regardless of the decoder used, a simplified binaural noising system is incorporated to account for human limitations in sound localization. As the distribution of sound source localization error is well understood [93, 94, 97–99], we apply Gaussian noise to the azimuth coordinates, $\tilde{\theta}_m$ (Equation 11), proportional to the error expected at that location, before passing them to the decoder. Effectively, the added noise drives the network to prefer central azimuth coordinates. As elevation is omitted from the encoding variables, we could fit a simple, exponential noising model to the localization error data points described in the study of Carlile, Leong and Hyams [97]: $\epsilon_m = 0.0647 * e^{\tilde{\theta}_m} - 0.0506$, shown in Figure 19 and 20.

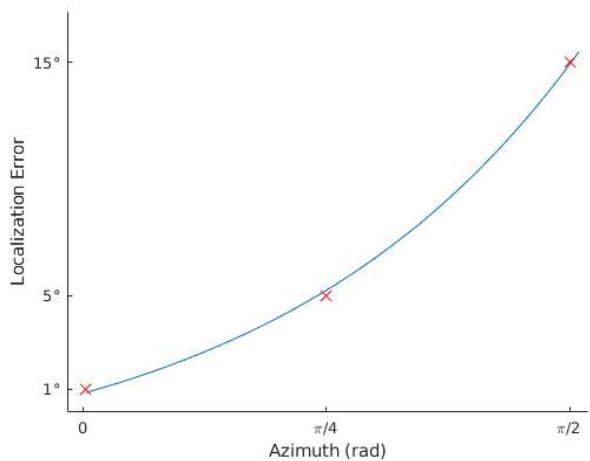

Figure 19: Simple fitted model of sound localization error by azimuth. The amount of noise imposed on $\tilde{\theta}_m$ is set to be proportional to the localization error.

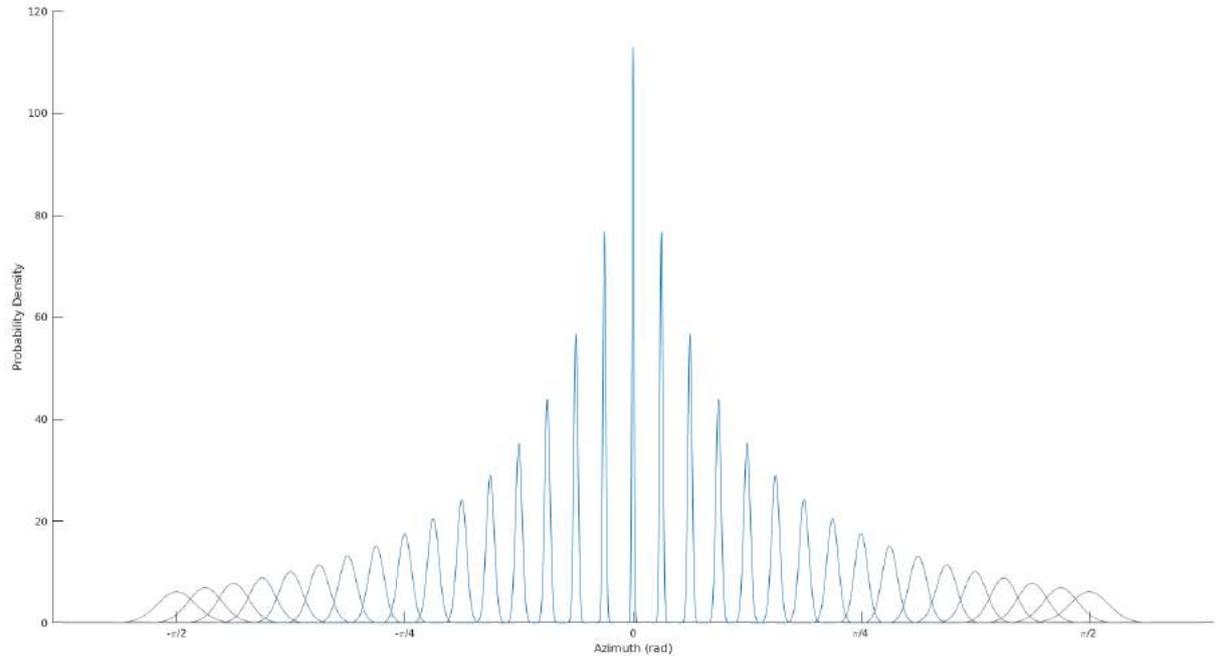

Figure 20: Probability density of Gaussian distributions of the applied binaural noising at various azimuth angles. The wider the distribution, the more the noise.

We have implemented four audio feature extraction networks, combinations of which are integrated in the hearing decoder. The MFCCs model [14] derives spectral information from the signal, but ignores phase. The discriminator network of WaveGAN [181] employs one-dimensional convolutions on the audio and likely retrieves phasic features. We also implemented a dilated causal convolutional network [104], which is effective to extract both spectral and phasic characteristics; however, it is still not autoregressive, meaning it merely realizes an FIR filter. Finally, we constructed a first-ever version of CARFAC [113] that can be embedded in DNNs, but due to its excessive amount of feedback loops, having a form of a complex IIR filter, it generates impractically huge unrolled network graphs, and trains extremely slow.

The hearing models rely on the following set of hyperparameters; typical values that were experimentally tested are shown as well:

- MFCCs frame length: window size, under which cepstral coefficients are computed; in literature, at least 24 ms is suggested to include shifts in low frequency signals as well; we found that setting it to the double of the section length yields better results

- MFCCs stride: time between beginnings of successive windows; set it as half the section length of the synthesized soundstreams

- Number of cepstral coefficients: we use all coefficients and expect the DNN to ignore the irrelevant ones over time; for speech recognition, 12 or 13 is recommended

- WaveGAN number of filters: convolutional filter count; we set it to 64

- WaveGAN kernel size: the temporal dimension of convolutional filters; similarly to MFCCs frame length, we set it to the double of the section length

- WaveGAN stride: stride of the convolutional filters; set it as half the section length

### 3.1.5 Reconstruction

Once the sound features are extracted and passed through the LSTM layers, the writer attention unit draws on the canvas given the decoded representation of the audio.

Two forms of drawing attention are proposed. First, the original grid pattern of Gaussian patches, same as applied in the reader attention. Second, a set of Gaussian patches distributed in a line instead of a grid, loosely resembling the visual feature space V1 simple cells code for. We termed the latter V1 attention.

V1 attention drawn edges are specified by their two-dimensional coordinates, orientation angle and the usual parameters regarding the Gaussian patches: isotropic variance and stride. The AEV2A model allows the drawing of multiple edges in a single iteration. The network is enforced to associate soundstream properties with lines, rather than arbitrary forms that grids of patches yield. Furthermore, our solution can individually link soundstreams to lines. Such a link provides the opportunity to enforce harmony between sound qualities of streams and the parameters of edges in the form of additional loss function terms. The model then converges to soundstreams corresponding to lines, incorporating research in audio-visual congruence: a higher pitch stream is associated with a line in the upper side of the canvas, and a sound source location on the right implies the related edge to be towards the right as well, i.e. spatial coherence. Thus, the drawback of implicit SS models is partially counteracted, and we can define audio-visual correspondence that the explicit approaches inherently possess.

We further specify a loss term encouraging higher audio amplitude to be bind to bigger visual objects, and higher "luminance" simultaneously. As we trained our models on contour images, luminance here refers to the strength of the contour. We achieve the amplitude–size and amplitude–luminance coherence, by guiding the amplitude of the soundstreams to be proportional to the amount of drawings made in a given iteration, essentially measuring the dissimilarity between the consecutive states of the canvas.

The reconstruction phase is guided by two hyperparameters. First, the number of Gaussian patches to apply both in the reading and writing attention, to which the network is barely sensitive. The second defines the number of edges to draw in one iteration of the network, in case we adopt the V1 attention. We set it to be equal to the amount of soundstreams we include in a soundscape, in order to obtain the benefits of audio-visual congruency described above.

### 3.1.6 Training

AEV2A was implemented in TensorFlow [186] and trained on NVIDIA's Tesla P100 GPU cards. Models installed with the hearing decoder required 2–3 GPUs and 1–2 days of training, while the ones integrating the deaf decoder ran on 1 GPU and converged within hours.

The Adam optimizer [187] was employed throughout. The learning was subjected to an exponential decay, starting between $10^{-4}$ and $5 \cdot 10^{-5}$, depending on the model configuration; hearing models were more likely to diverge and preferred lower learning rates. To further avoid divergence and exploding gradients, gradient clipping was introduced [188], along with batch normalization layers [189] placed after every dense neural network layer. Omitting these techniques resulted in divergence, or exploding gradients, according to our experiments.

The loss function we optimized on consisted of three terms: the reconstruction, latent and congruence losses. The reconstruction cost is defined as the L2 distance between the input image and the final state of the canvas. The latent loss is the sum of the Kullback–Leibler divergence of the latent prior we impose, from the distribution of our latent variables conditioned on the output of the encoder, summed over the iterations of the network [35]. Essentially, the distribution of our latent variables are sanctioned to follow a Normal distribution in every iteration. The congruence cost is separated into three components as mentioned in the previous section: the pitch–vertical, azimuth–horizontal and amplitude–size audio–visual congruence losses. The pitch–vertical loss is computed as the distance between the average pitch of the soundstreams and the vertical position of the corresponding visual feature, be that a grid or a line of Gaussian patches. The azimuth–horizontal cost is derived in a similar manner: it evokes coherence between the spatial position of the soundstream and the matching visual feature, on the horizontal plane. The amplitude–size or amplitude–luminance loss is defined for each iteration as follows: estimate the z-score of the amount of content drawn on the canvas, given the mean and standard deviation of added content in all iterations. Then take the mean amplitude of soundstreams in the same iteration, upscale it to the interval of $[-2, 2]$ to adjust for the ballpark of the previously calculated z-score. Finally, compute the squared difference between the two. This loss term is imperfect in the sense that it only encourages the amplitude–size correspondence between iterations of a single image; the same sized drawing on a different canvas may not have the same soundstream amplitude associated with it.

## 3.2 Case studies

We sought to test the efficacy of the AEV2E model in two case studies. First, datasets of images were generated to cater for the two distinct tasks. By performing these tasks, we aimed to investigate whether it is perceptually possible to associate the synthesized sounds to categories of images and whether one could execute accurate reaching movements relying solely on the audio encoded information. In the process, hundreds of AEV2E models were trained, before we deduced the proper set of hyperparameters and the overall architecture.

In the first case study, the author spent 5 days completely blindfolded, while occasionally performing SS training using one of the prepared AEV2E models. After 5 days, the accuracy to discriminate between images of different hand postures was tested.

In the second case study, the participant was blindfolded only during the training and testing periods. The task was to grab a beer can, or a gear shaped object sitting on a round table. This experiment was two-folded: first, the subject had to realize the object category, then reach for the object appropriately.

### 3.2.1 Datasets

A dataset of hand images was generated by first taking a video shot of a hand, displaying 15 different postures, at different horizontal and vertical positions. In the experiment, the participant only had to differentiate between 5 of these postures, shown in Figure 16. We extracted 10 frames a second from the video, rescaled the images to 160x120 pixels and applied the push-pull CORF model [71] to retrieve the contour.

Before training, the contour images were bundled into a dataset and separated into a 90% train and 10% test sets. We had 2720 images in total.

We recorded video shots of a round table, having either a beer can or a gear on top. In-between shots, one of the objects was moved around to cover most of the surface within the view of the camera. The videos recorded totaled at around 13 minutes. Similarly to the first study, 10 frames a second were extracted, resulting in 8160 images. To construct the dataset, the same rescaling and edge detection functions were applied as in the blindfolded case study.

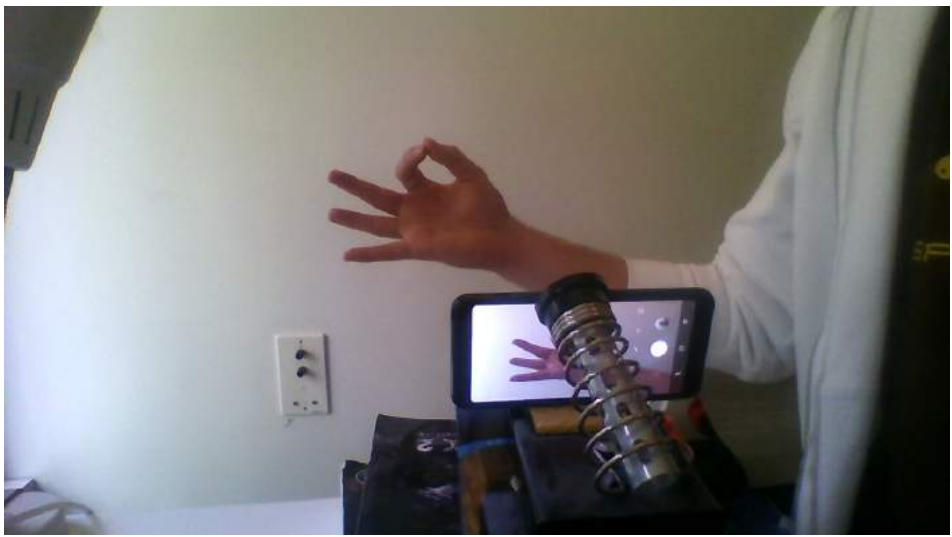

Figure 21: The minimalistic apparatus used to synthesize the dataset of hand postures.

### 3.2.2 Prototype

We had two objectives in mind when developing the AEV2A prototype: to keep the substitution delay at a minimum, and to restrict the view angle, leveling it to the eyes. Both of these objectives serve to hasten the SS training process. As discussed previously, the benefits of lower substitution delay builds on the study of perceptual learning, which says that temporally close sensory events are easier to associate. By restricting the view angle, we coupled the haptic and muscle feedback tighter to the sounds, as the user did not need to generalize over various angles. If we shifted the camera in every session, relatively to the direction and position of the head, the user would get confused over the incoherence of the soundscapes.

Both AEV2A models used in our experiments were trained with deaf decoders. We employed a V1 writer attention in the blindfolded case study, and grid attention patches in the reaching movement experiment. The parameter list for both models are shown in Table 2.

To real-time record images from a constrained view angle, we placed an Android phone inside a Google Cardboard. The phone was connected to a GPU-accelerated laptop via USB; video stream was broadcasted using an IP webcam application with USB tethering turned on. We achieved the lowest delay with such a setup, compared to sharing the video stream over WiFi or Bluetooth. The cardboard headset ensured that the camera view remained similar relative to head position and direction, across training sessions.

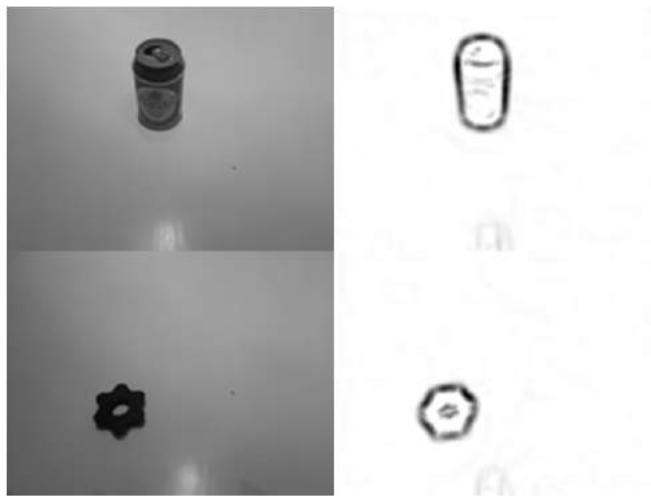

Figure 22: Example images and corresponding contour representations from the table dataset. A beer or a gear was placed in various positions on the table, while being recorded by a mounted mobile phone.

Table 2: Comparison of AEV2A model parameters used in the blindfolded, hand posture case study and in the reaching movement, table experiment. Due to the more complex forms present in the hand posture dataset, the corresponding autoencoder needed more than thrice the drawing iterations, which was further necessitated by the V1 writer attention unit: in general, V1 attention requires a longer sequence of recurrent iterations to be implemented. The congruence weight set to be an order of magnitude higher in the second experiment in order to enforce the relations between auditory and visual features of spatial location.

|                        | Hand posture | Reaching movement |
| ---------------------- | ------------ | ----------------- |
| Decoder type           | deaf         | deaf              |
| Attention type         | V1           | grid              |
| Congruence weight      | 0.1          | 1.0               |
| Sequence length        | 26           | 8                 |
| Number of soundstreams | 3            | 3                 |
| Number of sections     | 4            | 4                 |
| Section length (ms)    | 8            | 10                |
| Soundscape length (ms) | 998          | 480               |

We implemented a Python script, which ran indefinitely on the PC, grabbing frames from the video stream. The received images are rescaled, before their contours were extracted by the push-pull CORF model [71]. As we only had access to a Matlab implementation of CORF, we needed to establish a link between the main script and a Matlab session. We found pymatlab [190] to be the fastest approach in terms of the introduced delay, which remained under 150 ms. We could have reduced the latency by applying the Sobel operator instead of CORF, but we found CORF to be less noisy in practice. The AEV2A image-to-sound conversion function took approximately 100–200 ms to perform on single images, depending on the complexity of the model, particularly, on the number of iterations. We registered the moving average of the preceding delays, and grabbed the next image, so the corresponding soundscape would start just at the time the previous ended.

### 3.2.3 Experimental context

The author spent 5 days completely blindfolded, in order to attain superior hearing abilities, and to encourage the cross-modal recruitment of the occipital cortex [125]. He mostly consumed microwave Indian food, bananas and Skyr. For the majority of the 5 days, the author stayed in an apartment, leaving it solely on two occasions: once for a walk and once for a rave party. He experienced hallucinations from visual deprivation, and can I discern between flashing illusions of light and darkness. The author had visions of light blobs changing shapes, moving around. He saw different animals, such as wolves, snakes and eagles, mostly facing him. When blindfolded, he felt slightly depressed; he stayed in bed in the morning, and could not plan ahead further than a couple of hours. In the practical sense, living in the dark was easy, given that the author spent his days indoors and had friends helping him out. Otherwise, these 5 days were so gloomy, he decided against repeating it in the future.

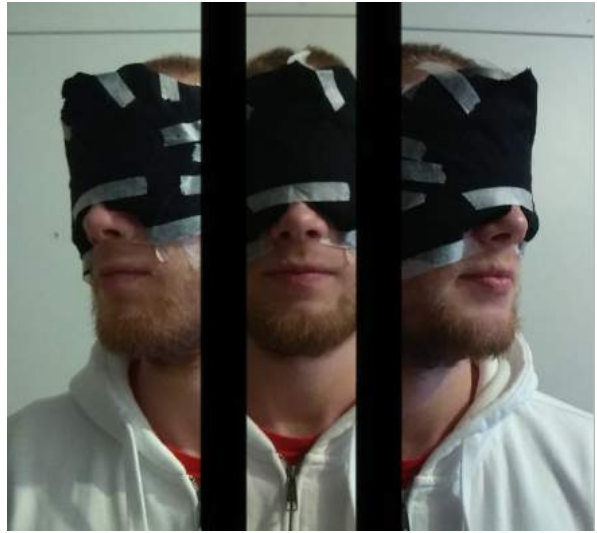

Figure 23: The author wearing the blindfold. The mask provided total visual abstinence.

In total, the author barely trained for 5 hours: more than 1 hour a day, except for the first day, when the session had to be skipped. Before going blindfolded, the AEV2A model used in the training process had not been tested or the conversion viewed, hence the subject had no visual intuition with regards to the audio–visual correspondence. This is critical, as the blind could neither obtain such an intuition.

Video stream of hand postures were real-time converted to soundscapes. To train myself, I held my hand in front of a white wall, wearing a cardboard VR headset, with my phone inside. While training, the camera was pointed towards my hand; during testing, it was recording someone else's hand. At each test case, I had to guess the hand posture, until the guess was right; the number of such guesses were recorded.

The purpose of the second case study was to determine whether the AEV2A model is reliable in spatial and shape representation of two object categories: a 0.33 liter beer can and a gear with a 3 cm radius. During training and testing sessions, one of the objects was placed on the round table having a radius of 0.5 m. The participant had to reach for, and pick up the object, while receiving the generated soundscapes in real time. The author had trained 2 hours in two separate occasions, before the testing session began.

Each grasping attempt was scored according to the method described in the study of Proulx and others [191]: an indirect movement was scored as 1, a relatively direct as 2. and a direct reach for the object as 3. Sweeping hand movements resulted in a score of 1, while constrained, fast search with the fingers at the right position was coded as 2. A score of 3 was attained when the hand posture was appropriate to grasp the object, and the reach was confident and accurate within a 3-cm radius of the center of the object. Knocking over the beer implied a score of 1, obviously.

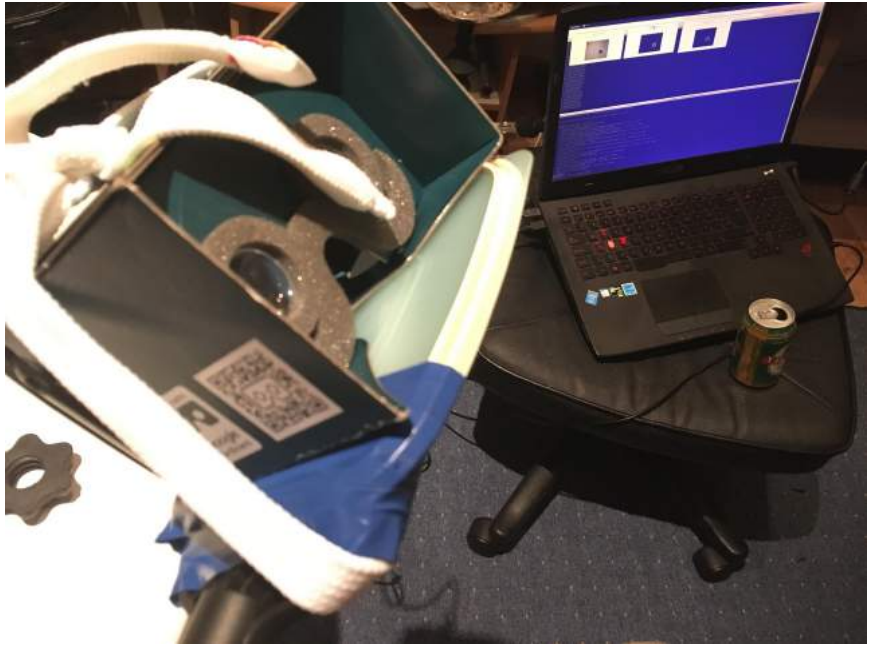

Figure 24: Experimental setup of the reaching movement case study. A Google Cardboard was custom attached to a tripod, which aimed the phone camera at a round table. Recorded images were real-time translated to sound on a GPU-accelerated laptop.

Similarly to the blindfolded experiment, the participant had not assessed visually the V2A conversion logic, and had not seen visual representation of the drawing iterations until the testing session ended. Such assessment would have built a visual intuition, which is inaccessible to the blind.

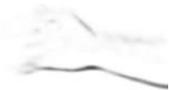

# 4 Results

## 4.1 Autoencoded Vision to Audition (AEV2A)

We trained more than 200 models, just on the hand posture dataset, to arrive at the architecture and configuration that we finally adopted in the case studies. At first, AEV2A instances suffered from instability and posterior collapse; we mitigated both by introducing batch normalization, gradient clipping and skip connections [184].

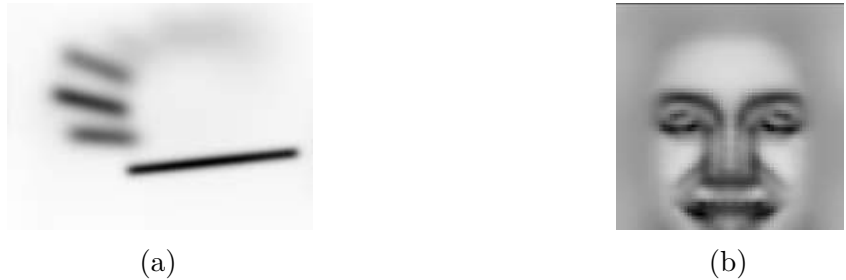

(a)            (b)

Figure 25: Examples of posterior collapse. AEV2A reconstructed images of the hand posture (a) and the CelebA [192] dataset (b). Essentially, they are the mean of their respective data samples, which the decoder learned to generate independent of the latent state.

We experimented with sequence length options between 6 and 42: the table dataset required much less iterations to converge, due to the simplicity of the dataset. Above 20 iterations, no improvements were gained (Figure 26), which could be due to the vanishing gradient problem. One of the models trained on the table dataset managed to draw the whole image in one iteration, while merely idling in the following five. Training time increases linearly with a slope of 1.57, as the sequence length rises.

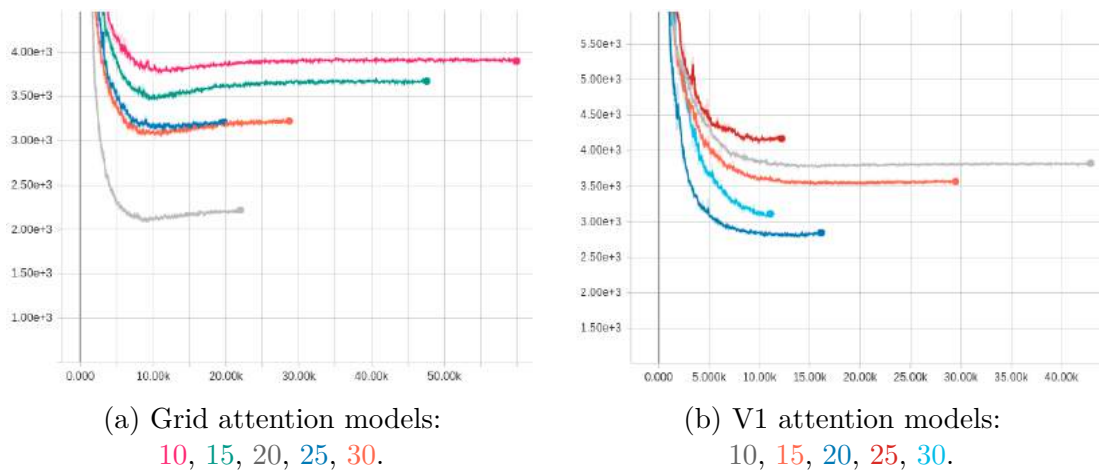

(a) Grid attention models:                 (b) V1 attention models:
10, 15, 20, 25, 30.                     10, 15, 20, 25, 30.

Figure 26: The effect of sequence length on the reconstruction loss computed on the test set. For each grid (a) and V1 (b) writer attention model, the sequence length is indicated by the coloring above. Note that the offsets of the Y axes are different.

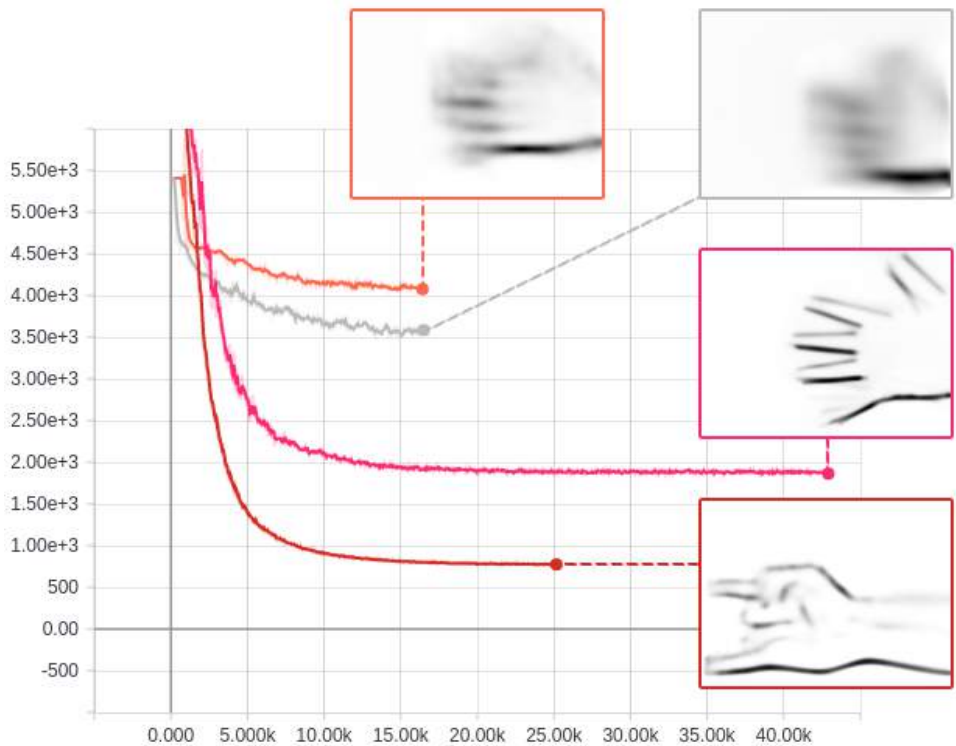

Figure 27: Reconstruction loss of the best performing models, comparing hearing (orange, gray) and deaf decoders (magenta, red). In general, hearing models fail to draw the fine details of fingers, and mostly learn to encode the position of the hand. A sample of the reconstructed test images is shown for each model.

Models equipped with hearing decoders were orders of magnitude slower to train and the reconstruction accuracy tended to plateau at suboptimal states (Figure 27). As of yet, we have not been able to configure a hearing model, which could represent the fingers properly on the reconstructed images of the posture dataset; we had models drawing blurry hands, at the right positions of the canvas. Overlapping soundscapes posed a serious obstacle for hearing decoders. In our experiments, we found the combination of MFCCs and WaveGAN models to perform the best at a relatively low computational cost, achieving only marginal improvements by including the dilated causal convolutional network.

The network optimized the congruence costs over time. The correspondence between the x-y coordinates of the visual object and the azimuth-pitch qualities of the sound were clearly perceivable. Further increasing the weight on the congruence loss term noticeably strengthened the symmetry between visual and auditory features (Figure 28). The amplitude–size correspondence was negligible.

Although one could perceptually distinguish between the audio representation of the two object types used in the reaching movement experiment, we performed a numerical assessment of the matter. The test dataset was manually labeled, the object present on each decoded image was identified. We compared two models trained on the reaching movement problem: the one we employed in the experiment (*A*), and another network with a larger bottleneck, but otherwise same hyperparameters ($B$).

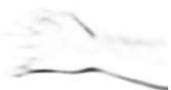

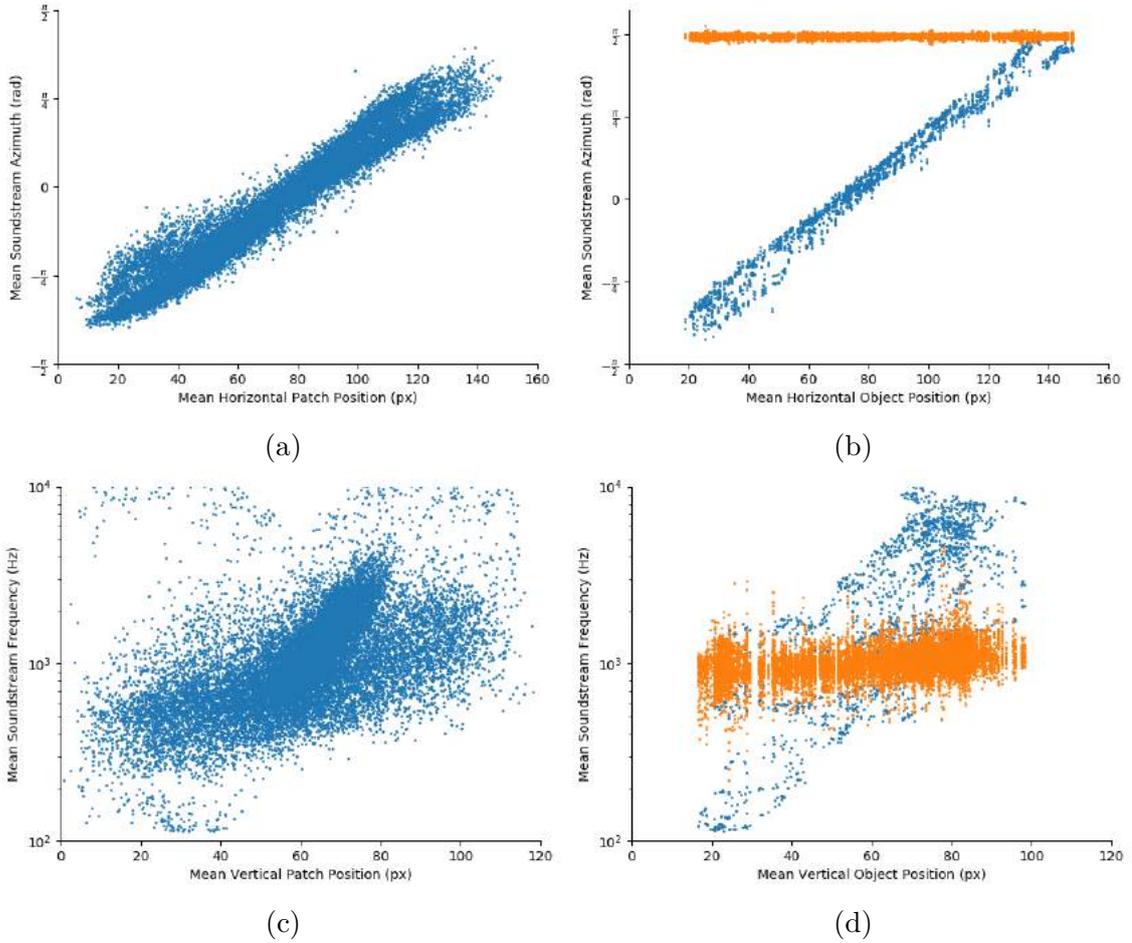

(a)                                                    (b)

(c)                                                    (d)

Figure 28: Relationship between decoded visual and corresponding audio features. The correlation is stronger for the network trained on the table dataset with a higher congruence weight (b, d), than the feature correspondence yielded by the hand posture sonifying model (a, c). Object position is computed as the geometrical mean of all drawings on the canvas, while patch position is derived from the V1 Gaussian patch parameters. The particular sound encoding of the table dataset trained network (b, d) required the separation of the first soundscape (blue) from the rest (orange), as only the former conveyed positional information, while the latter related mostly to object shape. Pearson correlation coefficients and p-values: (a) $r = .97$, $p < .01$; (b) $r = .99$, $p < .01$; (c) $r = .48$, $p < .01$; (d) $r = .90$, $p < .01$. All datapoints were generated by feeding the test set to the networks.

As can be seen in Figure 28, only the first soundscape described object position in case of network $A$: the first maintained a strong correlation with vertical and horizontal object location, while the rest of the soundscapes were indifferent. Model $A$ managed to further describe the shape in the first soundscape, though the rest of the audio was also representative of shape information. We ran k-means clustering ($k = 2$) on AM, FM and SM features of different subsets of the soundscape sequences in order to find those soundscapes that encoded object shape beside the first one. We compared the manual and cluster assigned labels, and the subset of features with the highest correlation was assumed to contain shape information. We found that the third soundscape alone suggested shape the most with a 62 % label equivalence.

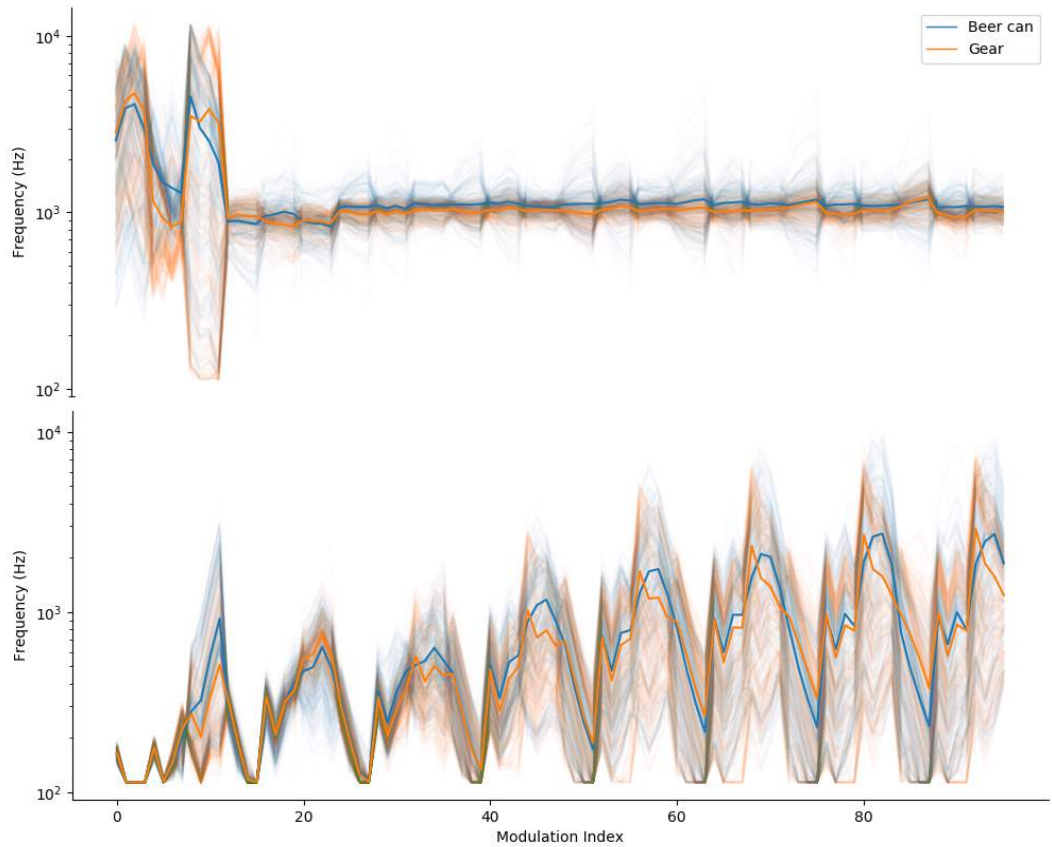

Figure 29: FM vectors across soundscapes of the beer can and gear audio representations. The semi-transparent lines are the modulation instances, the mean is depicted in opaque. Model *A* (top) encodes both position and shape in the first soundscape, while network *B* spreads those features more evenly across time.

Model *B* spread the shape encoding more evenly across soundscapes as the bottom row of Figure 29 suggests. In each soundscape, sonification of gear images attacked with a high frequency then dropped, beer can sounds had a slightly delayed frequency peak. We noticed similar dynamics in AM, and to some extent in SM, which indicates that shape was encoded in all these modulation vectors.

The binaural noising model successfully drove the model to exploit the central azimuth coordinates more than lateral locations, as shown in Figure 30. Due to the additional noising, the modulation values became more extreme, yielding more tenacious shifts in azimuth.

As for the comparison of writer attention units, grid drawing models consistently prevailed in reconstruction accuracy. V1 writer attention requires more drawing iterations in general. Furthermore, when equalizing the number of edges drawn and the amount of soundstreams in a soundscape, which is beneficial for enforcing audio-visual congruence, we found that the minimum quantity of modulations necessary is 4. Note that more modulations, that is, longer soundstreams, lead to a higher bandwidth of latent encoding allocated for each drawn edge.

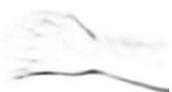

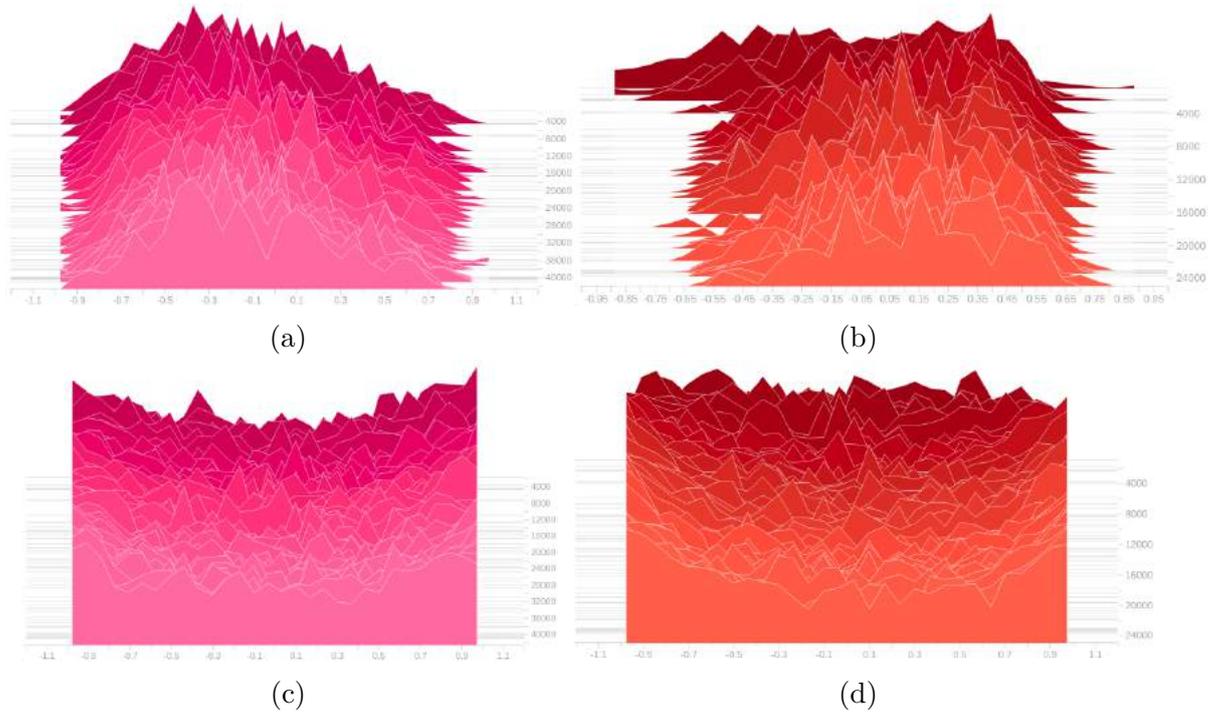

Figure 30: Example distributions of ground azimuth values (a, b) and modulation intensity (c, d). The depth dimension represents the number of training epochs, the X axis shows the azimuth value, ranging from –1 to 1 for both ground and modulation variables. Offset values converged towards central locations, while modulations became more extreme over training, as expected. Plots are from TensorBoard [193].

The thesis functions as a flip book. In the bottom left corner of every even page, we printed images depicting consecutive states of the canvas, following the sequence of drawings that a AEV2A model performs. Flip the pages to see the animated version of the sound-to-image reconstruction.

## 4.2   Case studies

For both case studies, the additional congruence costs successfully aided the subject to identify the position of the hand or the objects on the table; especially in the reaching movement experiment, for which the cost weighting was intentionally raised.

The AEV2A prototype suffered from substantial computation delay: 350 ms in the blindfolded case study, and 200 ms in the reaching movement experiment. The difference between the duration of delays was mostly due to the number of iterations, first model having 26, the second involving 6.

### 4.2.1 Hand posture

Learning to associate hand postures to soundscapes was found to be difficult. The difficulty was most likely due to the ever changing frequency, amplitude and azimuth ground values, as we failed to turn off the variational aspect of the autoencoder for the SS training and testing period. Because the Gaussian noise was still applied on these offset values, the same images induced soundscapes of slightly different spectral, intensity and spatial distributions. The ever shifting properties required the participant to generalize over them. Nevertheless, in average, the audio properties were informative enough to detail the position and posture of the displayed hand.

Furthermore, the applied AEV2A model emitted a high frequency "ting" sound for images, which included a hand or a hand-like object; for noisy input in particular, it did not. This indicator turned out to be extremely convenient during the training process, making the participant assured that the camera wielded the intended perspective.

A chi-square test of goodness-of-fit was performed to determine whether the number of hand posture guesses were equally distributed. The number of guesses were not uniformly distributed, being significantly less than the number of guesses at chance level, $\chi^2(4, N = 84) = 14.57$, $p < .01$. Hence, the subject was substantially more accurate in inferring the hand posture, aided by AEV2A encoded soundscapes, than random, brute-force guessing.

### 4.2.2 Reaching movement

To appropriately compare the added positional and categorical information that the AEV2A soundscapes provided, the participant underwent two tests: first, using the proper model trained on the table dataset; second, operating with a hand posture fed model. As the latter should have not conveyed sensible soundscapes in the context of this study, yet exerted comparable stimuli, it was ideal to demonstrate the attainable baseline, chance-level accuracy. However, the second model did provide scarce positional information, spatially playing sounds and drawing hands where the objects laid on the table, which only strengthens the statistical significance of the results. The reaching movement was performed within 10–15 seconds in average. The object identification accuracy amounted to 73%, compared to the 45% achieved listening to the baseline soundscapes.

An independent-samples t-test was conducted to compare the reaching movement accuracy in the two conditions. Results indicate a significantly better reaching movement accuracy for using the model trained on the table dataset ($M = 2.15, SD = 0.83$) over the application of the hand posture trained, baseline model ($M = 1.4, SD = 0.71$), $t(78) = 4.33$, $p < .0001$. Thus, the subject was able to identify the spatial properties of objects more accurately, compared to randomly reaching for objects on the table.

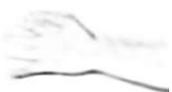

# 5 Discussion

There is an abundance of room to improve the AEV2A approach: the synthesized soundscapes are still longer than ideal, the image reconstruction reigns of instability, and the issue of human hearing compression is not addressed even close to its entirety. By improving on the architecture of the model, and by exploring environments and corresponding datasets in which this approach is appropriate, we should arrive at practical use cases. This thesis serves as a proof-of-concept for implicit conversion methods, and it demonstrates the potential they embody in the realm of vision rehabilitation. Two case studies were conducted, results of which indicate the shape and spatial encoding capabilities of the proposed model; though further experiments involving blind participants are definitely necessary to solidify the conclusions.

## 5.1 Autoencoded Vision to Audition

In terms of the training procedure, we encountered unstable results in the reconstruction accuracy and the capability for generalization. AEV2A instances trained on the same dataset with equivalent configuration yielded highly varying synthetization logic and reconstruction adequacy, which we attribute to the unpredictable effects of heavy gradient clipping in the process of convergence. To some extent, this issue can be alleviated by the introduction of deterministic warm-up [194] that guarantees increased training stability and smoother distribution of latent variables.

Beyond 20–30 recurrent iterations, AEV2A begins to degrade, most probably due to the vanishing gradient phenomenon and the uncertainty of repeated Gaussian noising. The latter can be addressed by substituting the variational latent space with a discrete representation [176]. A discrete bottleneck would further ensure a well-spread distribution of the soundscapes, and hence, perceptually distinguishable audio. Moreover, we could explicitly constrain the information content of the discrete space [178] to avoid the encoding of visual features in tiny, imperceivable shifts of sound that continuous variables may allow for [177]; such constraints may be informed by psychoacoustic research of human hearing limitations.

The additional congruence losses enforced the network to implement the pitch-vertical and azimuth-horizontal audio-visual correspondences. Further engineering of the cost function could render the audio representation more informative, by imposing additional tasks to learn, apart from reconstruction. In case of the hand posture encoding problem, the network could additionally infer hand keypoint positions, then the Euclidean distance between the real keypoints [195] and the predicted ones would serve as a loss term. Similar concept could be demonstrated to a AEV2A model, which is trained on images of an apartment: require the autoencoder to output the coordinates where the person stands in the apartment. Through such means, the network learns more about the visual objects, and could build a more coherent spatial representation of hands or indoor objects. On a side note, we attempted to train AEV2A instances on images of an apartment. However, due to the small sample set, which barely amounted to an 8-minute video, our model failed to converge properly, and the reconstruction cost remained high. Yet, the author was able to train himself to associate soundscapes to directions he looked towards, rotating on a chair, from a single position.

Although the audio synthesizer generates soundscapes that are, compared to other SS solutions, more pleasant to the ear, we foresee further development in this regard. An alternative approach is to employ a GAN architecture that learns to generate soundscapes given an arbitrary input vector [181]. After pre-training a GAN model to synthesize, e.g. Chopin pieces, we could incorporate the generator half of the network in AEV2A, receiving its input from the encoder unit, releasing Chopin excerpts in turn to the decoder. Consequently, the generated audio may become pleasant and tailored to one's taste. However, there is no guarantee that different input parameters would result in perceptually distinct soundscapes. Furthermore, the output of such a GAN generator is strictly audio, hence, we would need to resort to hearing decoders. Traditional, explicit conversion methods, like TV and EyeMusic, need to insert a cue sound before playing the next soundscape, so the user can synchronize with the predefined scanning process. However, we found that AEV2A does not require such cue sounds to be injected, as the sonification process is not symmetric, that is, the beginning and the end of the soundscapes are distinctive enough to be perceived.

The proposed AEV2A model fails to exploit harmonics and timbre coding in general. Even though soundstreams can be overlapped to sonify sounds of various timbre, we cannot be sure that such soundscapes are perceptually distinguishable. Moreover, to completely cover the space of perceivable timbre, we would need to support arbitrary patterns of overlap by defining and optimizing on the time offset of soundstreams within a soundscape; i.e. a differentiable variable is necessary, which determines the delay of streams. Previous SS devices introduced musical instruments as discrete agents of harmonics. Similarly, instruments may be incorporated into AEV2A models with deaf decoders, each instrument having its own AM, FM and SM states.

In order to render the SS learning curve smoother for the blind, the auditory-visual correspondence of the trained model should be uncovered as much as possible. By giving an explicit explanation of the influence that sound features elicit on the drawing primitives, we may ease the initial period of the learning process. Drawing primitives may be the position or angle of the patch printed on the canvas. If we enjoy the luxury of (partially) labeled datasets, we can play random sound sequences associated with one type of drawing or object; manual switching between labels in such a way could build an intuition on the distribution of soundscapes conditioned on the object type, disregarding the location information. Similarly, we can slightly shift either the synthesized sound qualities or the resulting drawing properties and present the corresponding visual or auditory features, respectively. Further intuition may be deduced from generating plots similar to Figure 29, visualizing the sound encoding of object types across modulations of frequency, amplitude or azimuth.

In our experiments, AEV2A instances installed with hearing decoders converged to a suboptimal state, and failed to reconstruct details of images, like the fingers and their orientations (Figure 27). Deaf decoders performed substantially better in comparison. Hearing decoders would definitely improve by substituting the combination of MFCCs and WaveGAN networks with a WaveNet [115] unit. Furthermore, we would see two major benefits from implementing computational models of human hearing in a deep learning environment: first, they function as effective auditory feature extractors, second, they would constrain the synthesizer to produce humanly perceivable soundwaves. We implemented a TensorFlow version of CARFAC [113], but it requires efforts of massive optimization until becoming practical to fuse with any deep learning solution.

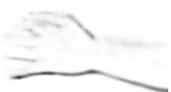

Regarding computational hearing models and SS devices, a promising experiment would be to examine whether explicit conversion functions construct audio that is subject to undesirable compression along the auditory pathway. In such an experiment, the hearing model could stay outside of the deep learning environment. The testing setup could be built as follows: take a diverse set of images, convert them to sound using an explicit SS device, pass the sound through a computational hearing model, take the output and train a deep learning model to associate it with the original image. This test might indicate ways to improve explicit SS devices, or may point out the set of images they struggle to represent reliably.

In terms of deaf decoder models, we may consider simple noising functions, which would penalize the plenitude of overlapping streams, according to psychoacoustic and neuroimaging studies of audio stream perception. Similarly to the binaural noising mechanism employed in this study, Gaussian noising layers could be designed, which would apply uncertainty to other audio qualities, proportional to the jnd associated to them. Finally, by including the elevation dimension of sound localization, coupled with a more complex binaural noising model on the hearing end, we could further expand the latent space.

The attention mechanism may also be subject to reformulation. The V1 writer attention proved more intuitive to learn, compared to the grid writer, as we found easier to associate lines with soundstreams, rather than arbitrary drawing of grid patches. However, V1 attention models yielded worse reconstruction accuracy in general, and required more iterations to generalize well. To explore more options, reading and writing units could be exchanged for convolutional and transposed convolutional networks. Some of the AEV2A models tend to spend the first couple iterations only reading and barely drawing; the network first needs to scan the image. This initial scanning might be accelerated by inserting a convolutional stage before the first iteration to extract relevant visual features of the whole image in one pass. Finally, we established that shade or color information can hardly be learned perceptually by the blind, however, we could test whether a dense visual representations, like grayscale images, facilitate the training of the autoencoder.

## 5.2 Applications

Selecting the right dataset is essential for training deep learning methods, and AEV2A is no different. In order to find practical use cases for the proposed model, small, contained environments should be explored. By recording videos of such environments, we could generate datasets with ease. As suggested, guiding the visually impaired in their apartments, or on streets that cause difficulties to stroll through may lead to functional solutions, even with the current prototype. In domains of low visual variation, where fast reaction time is necessary, AEV2A could adopt a short substitution delay, encoding only relevant features into audio, instead of the whole image as explicit methods do.

Classic, simple video games that only the sighted has enjoyed so far, could be made available for the blind. Atari and Nokia games like Asteroids, Night Driver, Snake and Space Impact require rapid control from the user, and manifest simple enough, low-variance visual features, which an implicit conversion logic can exploit, resulting in lower substitution delays. By taking representative screenshots of these games, we can generate datasets that a AEV2A model can train on. As semantic segmentation [196] simplifies the visual scene for self-driving cars by extracting actionable features, such segmentation can be fed as an input to an implicit V2A conversion method to represent the lower complexity imagery in soundscapes for the blind.

In certain cases, in which the environment and the task to accomplish within are simple enough, one could manually design the V2A conversion; in other circumstances where such a hand-crafted design becomes nontrivial, AEV2A is applicable. For instance, developing a substitution method for bullseye shooting is straightforward: convert the distance between the bullseye and the point of aim to the frequency domain, so higher frequency would translate to a more accurate aim. Although, such a conversion requires extra components to be attached to the gun, the audio representation is easy to learn. However, the audio translation of various human and animal shaped targets is much less clear and may necessitate the employment of explicit or implicit conversion techniques; and if such targets are moving, an implicit solution is most likely essential.

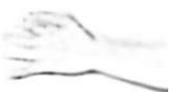

# 6    Conclusion

This thesis systematically reviewed the literature of V2A SS and relevant fields of cross-modality, sensory coding, psychoacoustics, perceptual and deep learning. The significance of human hearing limitations and the cross-modal recruitment of the visual cortex were emphasized, while SS solutions were presented that comply to the limitations and aspires to exploit the recruitment.

This study highlighted the influence of substitution delay and human hearing limitations on the struggle of SS training, and revealed approaches to reduce the delay, including visual space abstraction, i.e. contour extraction, and implicit conversion methods. In these regards, we enumerated the available edge detection algorithms and developed a deep recurrent VAE to perform image-to-sound translation.

The designed DNN contains a hand-crafted sound synthesizer that incorporates human hearing limitations, and additionally, the network is able to accommodate computational hearing models to further manifest such limitations; a binaural noising unit was successfully employed, so central azimuth values were exploited more likely than lateral ones, proportional to our spatial audio localization accuracy. We trained more than 200 AEV2A networks, before we arrived at two models that we finally tested separately in blindfolded experiments.

One case study examined the applicability of AEV2A in categorical shape discrimination, while the second test investigated whether spatial properties of objects are reliably encoded in this V2A SS scheme. A few hours of training yielded significantly better performance than baseline, which demonstrated the viability of AEV2A to offer a rapid SS learning rate.

We envision implicit SS solutions to be specifically trained for various environments, providing the opportunity for the blind to work and play where visual context is a must, a tighter grip on the substitution delay is necessary, while a guarantee on lossless hearing is paramount.

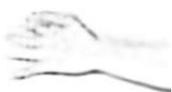

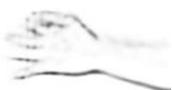

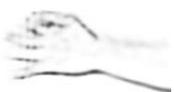

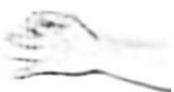

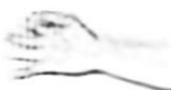

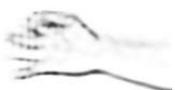

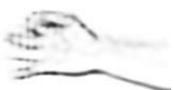

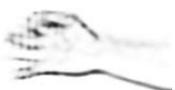

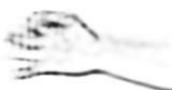